\documentclass[letterpaper,twocolumn,showpacs,preprintnumbers,amsmath,amssymb,floatfix,superscriptaddress,pre]{revtex4}

\usepackage{psfrag,color}
\usepackage{graphicx}  % Include figure files
\usepackage{dcolumn}   % Align table columns on decimal point
\usepackage{bm}        % bold math
\usepackage{epsfig}
\usepackage{textcomp}
\usepackage{subfigure}
\usepackage[sort&compress]{natbib}
\definecolor{rred}{rgb}{.7,0,0}

\newcommand{\secn}[1]{\section{#1}}

\newcommand{\subsecn}[1]{\subsection{#1}}
\newcommand{\subsubsecn}[1]{\subsubsection{#1}}
\newcommand{\subsecnnn}[1]{\subsection*{#1}}
\begin{document}

\title{Universality of modulation length (and time) exponents}

\author{Saurish Chakrabarty}
\affiliation{Department of Physics and Center for Materials Innovation, Washington University in St Louis, MO 63130, USA.}

\author{Vladimir Dobrosavljevi\ifmmode \acute{c}\else \'{c}\fi{}}
\affiliation{Department of Physics and National High Magnetic Field Laboratory, Florida State University, Tallahassee, Florida 32306, USA.}

\author{Alexander Seidel}
\affiliation{Department of Physics and Center for Materials Innovation, Washington University in St Louis, MO 63130, USA.}

\author{Zohar Nussinov}
\email{zohar@wuphys.wustl.edu}
\affiliation{Department of Physics and Center for Materials Innovation, Washington University in St Louis, MO 63130, USA.}
\affiliation{Kavli Institute for Theoretical Physics, Santa Barbara, CA 93106.}

\date{\today}

\begin{abstract}
We study systems with a crossover parameter $\lambda$, such as the
temperature $T$, which has a threshold value $\lambda_*$ across which the correlation function changes from
exhibiting fixed  wavelength (or time period) modulations to
continuously varying modulation lengths (or times). 
We report on a {\em new exponent}, $\nu_L$, characterizing the universal nature of this crossover.
These exponents, similar to standard correlation length exponents,
are obtained from motion of the poles of the momentum (or frequency) space
correlation functions in the complex $k$-plane (or $\omega$-plane) as
the parameter $\lambda$ is varied.
Near the crossover (i.e., for $\lambda\to\lambda_*$), the characteristic modulation wave-vector $K_R$ on
the variable modulation length ``phase'' is related to
that on the fixed modulation length side, $q$ via $|K_R-q|\propto|T-T_*|^{\nu_L}$.
We find, in general, that $\nu_L=1/2$. In some special instances, $\nu_L$ may attain other rational values.
We extend this result to general problems in which the eigenvalue of an operator
or a pole characterizing general response functions may attain a constant real (or imaginary) 
part beyond a particular threshold value, $\lambda_*$.
We discuss extensions of this result to multiple other arenas.
These include the axial next nearest neighbor Ising (ANNNI) model.
By extending our considerations, we comment
on relations pertaining not only to the modulation lengths (or times) but
also to the standard correlation lengths (or times).
We introduce the notion of a Josephson timescale.
We comment on the presence of aperiodic ``chaotic'' modulations in ``soft-spin''
and other systems.
These relate to glass type features.
We discuss applications to
Fermi systems -- with particular application to metal to band
insulator transitions, change of Fermi surface topology, divergent
effective masses, Dirac systems, and topological insulators.
Both regular periodic and glassy (and spatially chaotic behavior) may be found in strongly correlated electronic systems.
\end{abstract}

%Ising model - lattice theory, 05.50.+q
%Magnetic ordering - classical spin models, 75.10.Hk
%Magnetic ordering - general theory and models of, 75.10.-b
%High-temperature techniques and instrumentation, 07.20.Ka
%Correlations - collective effects, 71.45.Gm
%Magnetic ordering - general theory and models of, 75.10.-b
%Magnetic ordering - quantized spin models, 75.10.Jm
%Magnetic domains, 75.60.Ch
%Strongly correlated electron systems, 71.27.+a

\maketitle

%%%%%%%%%%%%%%%%%%%%%%%%%
\secn{Introduction}
%%%%%%%%%%%%%%%%%%%%%%%%%

\begin{figure}
  \begin{center}
\includegraphics[width=\columnwidth]{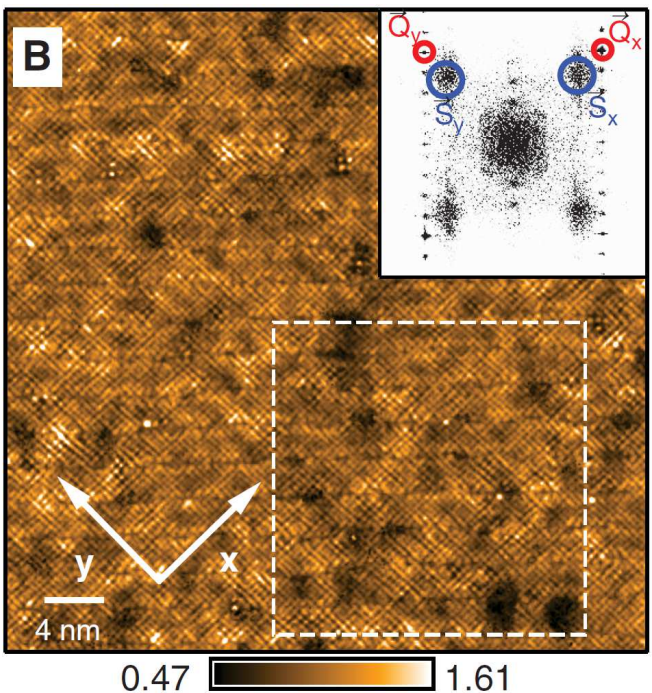}
    \caption{
Sub-unit-cell resolution image of the electronic structure of a
cuprate superconductor at the pseudo-gap energy. 
Inset shows Fourier space image of the same figure.
Nematic and smectic phases are highlighted using the red and blue
circles respectively.
The nematic phase is characterized by commensurate wave-vectors $\vec{Q}$.
The smectic wave-vector, on the other hand takes incommensurate
values, $\vec{S}$ which is dependent on the amount of doping, albeit weakly.
(From Ref. \cite{mesaros}.
Reprinted with permission from AAAS.)
}
    \label{mesaros1b}
  \end{center}
\end{figure}

In complex systems, there are, in general, possibly many important
length and time scales that characterize correlations. Aside from
correlation lengths describing the exponential decay of correlations,
in some materials there are length scales that characterize periodic
spatial modulations or other spatially non-uniform properties
as in Fig. \ref{mesaros1b}.
We investigate the evolution of these
length scales as a function of some parameter $\lambda$. 
This parameter may be the temperature, the chemical potential, or
some other physical quantity relevant for description of the system
being studied. To illustrate our basic premise, we will largely focus
on temperature dependences of the correlation function in this
work. However, with a trivial change of variables, our results are valid for any
parameter that, when tuned, connects a phase with continuously
varying modulation lengths (or times) to one in which the modulation
length (or time) is pinned to a fixed value.
The crossovers we consider are not symmetry breaking transitions. Consequences of our considerations also
relate to correlation lengths as we will comment on later.

Many systems exhibit subtle changes in their correlation functions at certain special temperatures.
We focus on situations wherein as the temperature is varied across a certain crossover temperature $T_*$, an unmodulated phase of a system (appearing  at $T>T_*$ or at  $T<T_*$) may start exhibiting
modulations (at $T<T_*$ or at $T>T_*$ respectively) sans a thermodynamic phase transition at $T_*$.
A generalization of this occurs when modulations in a system 
are characterized by a fixed wavelength on one side of 
a crossover temperature and by continuously varying wavelengths on the
other side.
Such an occurrence may generally rear its head when
interactions of different scales compete with one another.
A wealth of interesting periodic spatial patters appear arenas:  e.g., the manganites,\cite{salamonmanganites}
pnictide \cite{pnic1, pnic2} and
cuprate \cite{tran, cuprate1, cuprate2,
  cuprate3, cuprate4, steve} superconductors,
quantum Hall systems,\cite{qhstr1, qhstr2, qhstr3}
dense nuclear matter,\cite{pasta1, pasta2}
magnetic systems,\cite{magnstr1, magnstr2, magnstr3, magnstr4,
  magnstr5, magnstr6}
heavy fermion compounds,\cite{jan,rmp_steve}
membranes,\cite{membranestripes}
cholesterols,\cite{cholesterolstripes}
magnetic garnets,\cite{magneticgarnet} 
dipolar systems,\cite{dipole1,vindigni}
systems with nematic phases,\cite{nematicma}
and countless other systems.\cite{Seul,liebfrusferr,ortix,gulacsi,barci}

\secn{Our main results and their implications}
In this work, we report on the temperature (or other parameter) dependence of
emergent modulation lengths that govern the size of 
various domains present in some systems. 
In its simplest incarnation, our central result is that
if fixed wavelength modulations characterized by a particular {\em finite} length scale, $L_*$,
appear beyond some temperature $T_*$ 
then, the modulation length, $L_D$ on the other side of the crossover
differs from $L_*$ as
\begin{eqnarray}
|L_D-L_*| \propto |T-T_*|^{\nu_{L}}.
\end{eqnarray} 
When there are {\em no modulations} on one side of $T_*$, i.e.,
$L_*\to\infty$,
we have near the crossover,
\begin{eqnarray}
L_D\propto|T-T_*|^{-\nu_L}.
\end{eqnarray}
Apart from some special situations, we find that
irrespective of the interaction, $\nu_{L} = 1/2$.
We arrive at this rather universal result 
assuming that there is no phase transition at the crossover temperature $T_*$.
Our result holds everywhere inside a given thermodynamic phase of a system.

Our considerations are not limited to continuous crossovers. A corollary of our analysis pertains to systems with discontinuous ({\em
``first-order'' like}) jumps in the correlation or modulation lengths.

We will further comment on situations in which wherein a branch point appears at $T_*$. 
We will present examples where we obtain
rational and irrational exponents and also the anomalous critical
exponent $\eta$. Our analysis affords general connections to the 
critical scaling of correlation lengths in critical phenomena.

Our results for length scales can be extended to timescales. We will, amongst other notions, in employing a formal interchange of spatial with temporal coordinates,
introduce the concept of a {\it Josephson timescale}.

Lastly, further deepening the analogy between results in the spatial and time domain, we will comment on the presence of phases with
aperiodic spatial ``chaotic'' modulations (characteristic of amorphous configurations) in
systems governed by non-linear Euler-Lagrange equations. Aperiodic
``chaotic'' modulations may appear in strongly correlated electronic systems.

In the appendix, we present applications to Fermi systems pertaining 
to metal--band insulator transition, change of Fermi surface topology,
divergence of effective masses, Dirac systems and topological insulators.

%%%%%%%%%%%%%%%%%%%%%%%%%%%%%%%
\secn{The systems of study}
\label{systems_of_study}
%%%%%%%%%%%%%%%%%%%%%%%%%%%%%%%
In this work, we will predominantly consider translationally invariant systems on a lattice
whose Hamiltonian is given by 
\begin{eqnarray}
H = \frac{1}{2} \sum_{\vec{x}\neq\vec{y}}V(|\vec{x}-\vec{y}|)S(\vec{x}) S(\vec{y}). 
\label{Ham}
\end{eqnarray} 
The quantities $\{S(\vec{x})\}$ portray classical scalar spins or
fields. The sites $\vec{x}$ and $\vec{y}$ lie on a $d$-dimensional hyper-cubic (or
some other) lattice with $N$ sites. We will set the lattice constant to unity. [In the quantum arena, we
replace the spins $\vec{S}(\vec{x})$ in Eq. (\ref{Ham}) by Fermi or
Bose or quantum spin operators.]

The results that will be derived in this work apply to a variety of
systems. These include theories with trivial $n$-component generalizations of Eq. (\ref{Ham}). 
In the bulk of this work, the Hamiltonian has a bilinear form in the spins. We will however, later on,
study ``soft'' spin model with explicit finite quartic terms as we now expand on. 
% , wherein the ``softness'' is
% incorporated by a local quartic term in the Hamiltonian. For example,
An $n$-component generalization of Eq. (\ref{Ham}) is given by the Hamiltonian 
\begin{eqnarray}
H &=& \frac{1}{2}
\sum_{\vec{x}\neq\vec{y}}V(|\vec{x}-\vec{y}|)\vec{S}(\vec{x}) \cdot
\vec{S}(\vec{y})+\nonumber\\
&&\frac{u}{4}\sum_{\vec{x}} \left(\vec{S}(\vec{x})\cdot\vec{S}(\vec{x})-n\right)^2.
\label{Ham_soft}
\end{eqnarray} 
Such a Hamiltonian represents standard (or ``hard'') spin or $O(n)$ systems if $u\gg1$ in the large $u$ limit,
the quartic term enforces a ``hard'' normalization constraint of the particular form $\vec{S}(\vec{x})\cdot\vec{S}(\vec{x})=n$.
For finite (or small) $u$, Equation (\ref{Ham_soft}) describes
``soft''-spin systems wherein the normalization constraint is not
strictly enforced.
% Effects of such higher order terms will be discussed later.

% We will also discuss the evolution of general correlation functions in systems in
% which the exact form of the Hamiltonian might not be known.

% Even though in the bulk of this talk we discuss systems described by
% classical scalar spins, we also discuss applications to
% multi-component order parameters and also to quantum spins/fields.

In what follows, $v(k)$
and $s(\vec{k})$ will denote the Fourier transforms of
$V(|\vec{x}-\vec{y}|)$ and $S(\vec{x})$.
We employ the following Fourier conventions,
\begin{eqnarray}
a(\vec{k})&=&\sum_{\vec{x}}A(\vec{x})e^{i\vec{k}\cdot\vec{x}},\nonumber\\
A(\vec{x})&=&\frac{1}{N}\sum_{\vec{k}}a(\vec{k})e^{-i\vec{k}\cdot\vec{x}}.
\end{eqnarray}
With these conventions in tow, in Fourier space, Eq. (\ref{Ham}) reads
\begin{eqnarray}
H=\frac{1}{2N}\sum_{\vec{k}}v(k)|s(\vec{k})|^2.
\label{Hamkspace}
\end{eqnarray}
% For analytic interactions,
% $v(k)$ is a function of $k^{2}$ (to avoid branch cuts).
When $v(\vec{k})$ is analytic in all momentum space coordinates, it is a function of $|\vec{k}|^{2} = k^2$ (and not a
general function of  $k \equiv \sqrt{\sum_{l=1}^d k_l^2}$ with $\{k_l\}$ being the Cartesian components of $\vec{k}$).
This is so as $|\vec{k}|$ has branch cuts when viewed as a function of a particular $k_{l}$ (with all other $k_{l' \neq l}$ held fixed).  
The lattice Laplacian that links nearest neighbors sites in real space becomes 
\begin{eqnarray}
\label{LLaplace}
\Delta_{\vec{k}} = 2 (\sum_{l=1}^{d} (1- \cos k_{l})
\end{eqnarray}
in $k$-space. $\Delta_{\vec{k}}$
veers towards $|\vec{k}|^{2}$ in the continuum (small $k$) limit.
The two point correlation function for the system in Eq. (\ref{Ham})
is,
$ %\begin{eqnarray}
G(\vec{x})=\langle S(0)S(\vec{x}) \rangle.
$ %\label{Gx} \end{eqnarray}
At large distances $x=|\vec{x}|$, the correlation function has a general asymptotic behavior
\begin{eqnarray}
G(x)\approx\sum_i f_i(x)\cos\left(\frac{2\pi x}{L_D^{(i)}}\right)e^{-x/\xi_i}.
\label{gxtyp}
\end{eqnarray}
In the $i$-th term, $f_i(x)$ is an algebraic prefactor, 
 $L_D^{(i)}$ is the modulation length
and $\xi_i$ is the corresponding correlation length.
In general, the function $f_i(x)$ may contain a factor with an anomalous exponent $\eta$ (usually not an integer), such as,
$f_i(x)\propto1/x^{d-2+\eta}$.
% Most of our results are derived assuming the absence of such behavior,
% i.e., $\eta=0$.
% However, we will comment later about situations with non-zero $\eta$.
Generally, there can be multiple correlation and modulation lengths.
In Fourier space, 
%\begin{eqnarray}
$G(\vec{k})=\frac{1}{N}\langle |s(\vec{k})|^2 \rangle.$
%\end{eqnarray}
The modulation and correlation lengths can be obtained respectively from the real and imaginary parts of the
poles of $G(\vec{k})$ in the complex $k$-plane.

%%%%%%%%%%%%%%%%%%%%%%%%%%%%%%%%%%%%%%%%%%%%%%%%%%
\subsecnnn{General considerations: Correlation and modulation lengths from momentum space
  correlation function}
%%%%%%%%%%%%%%%%%%%%%%%%%%%%%%%%%%%%%%%%%%%%%%%%%%
The correlation function $G(\vec{x})$ in ($d$-dimensional) real space is related to
the momentum space correlation function $G(\vec{k})$ by
\begin{eqnarray}
G(\vec{x})=
%{\cal P}
\int\frac{d^dk}{(2\pi)^d} G(\vec{k}) e^{-i\vec{k}\cdot\vec{x}}.
\label{gkgxint}
\end{eqnarray}
% where ${\cal P}$ stands for the principal part of the integral.
On the lattice, the integral above must be replaced by summation over $\vec{k}$-values
belonging to the first Brillouin zone. In the continuum, which we discuss here, the integral
range is unbounded.
Even in lattice systems, doing an unbounded summation  over
$\vec{k}$-values provides a good approximation for the correlation
function in real space in many scenarios.

For spherically symmetric problems, i.e., when $G(\vec{k})=G(k)$,
\begin{eqnarray}
G(x)&=& % {\cal P} 
\int_0^\infty \frac{ k^{d-1}
  dk}{(2\pi)^{d/2}}\frac{\mbox{J}_{d/2-1}(kx)}{(kx)^{d/2-1}} G(k),
% \\ &=&{\cal P} \int_{-\infty}^\infty \frac{2 k^{d-1}
%   dk}{(2\pi)^{d/2}}\frac{\mbox{J}_{d/2-1}(kr)}{(kr)^{d/2-1}} G(k) 
\label{gksphsym}
\end{eqnarray}
where J$_\nu(x)$ is a Bessel function of order $\nu$.
The above integral can be evaluated by choosing an appropriate contour
in the complex $k$-plane. The contour can be closed along a circular
arc of radius $R\to\infty$ provided
\begin{eqnarray}
% |G(k)|\lesssim \frac{1}{k^{\frac{d+1}{2}}},\mbox{ as } k\to\infty.
|G(k)|\lesssim k^{-\frac{d+1}{2}},\mbox{ as } k\to\infty.
\end{eqnarray}
In evaluating the integral in Eq. (\ref{gksphsym}), we obtain contributions
from residues associated with the poles of the integrand as well as
contributions from its branch points. We use $K=K_R+iK_I$ to represent the poles and
branch points of the integrand in the complex plane.
The correlation and modulation lengths in the system are determined
respectively by the imaginary ($K_I$) and real parts ($K_R$) of these
poles and branch points.
Together, all these singularities can be compactly expressed as
\begin{eqnarray}
\boxed{\frac{1}{G^{(m)}(K)}=0},
\label{lengthsK}
\end{eqnarray}
where $0\le m<\infty$ is the order of the smallest
order derivative of $G(k)$ which diverges at $k=K$.\cite{mzerononzero}

% The arguments made here hold for systems
% defined in the continuum. Extensions to lattice systems is
% straightforward.
In footnote \cite{noteentire}, we comment on the situation in which the function $G(T,k)$ is an
entire function of $k$ (i.e., when is $G$ is analytic everywhere).

%%%%%%%%%%%%%%%%%%%%%%%%%%%%%%%%%%%%%%%%%%%%%%%%
\secn{A universal domain length exponent -- Details of analysis}
\label{universal_section}
%%%%%%%%%%%%%%%%%%%%%%%%%%%%%%%%%%%%%%%%%%%%%%%%
We now derive (via various inter-related approaches), our central
result -- the existence of  
a new exponent for the domain length
in rather general systems with
% Our result below applies to {\em general fields} [i.e., 
real or complex scalar fields, vectorial (or tensorial) fields of both the discrete 
(e.g., Potts like) and continuous variants.

\begin{figure*}
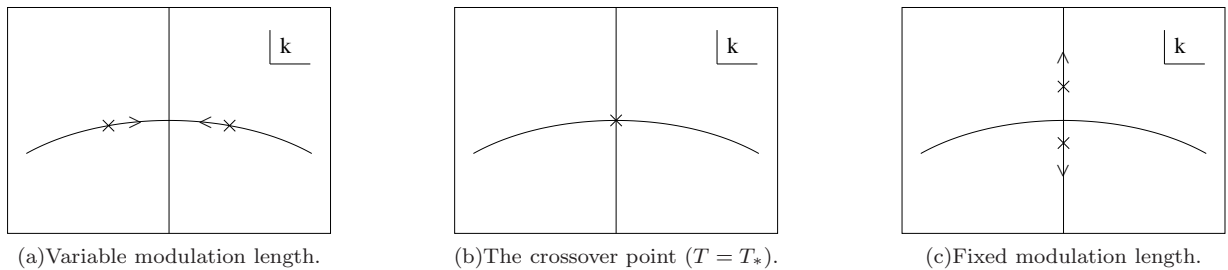

  \begin{center}
\subfigure[ Variable modulation length.]%[$T<T_*$]
{\includegraphics[width=.5\columnwidth]{cixpolesarrowsbefore.eps}}\qquad
\hspace{.1\columnwidth}
\subfigure[ The crossover point ($T=T_*$).]{\includegraphics[width=.5\columnwidth]{cixpolesarrowststar.eps}}\qquad
\hspace{.1\columnwidth}
\subfigure[ Fixed modulation length.]%[$T>T_*$]
{\includegraphics[width=.5\columnwidth]{cixpolesarrowsafter.eps}}
    \caption{Schematic showing the trajectories of the singularities
      of the correlation function near a fixed -- variable modulation length
      crossover. Two poles of the correlation function merge at
      $k=k_*$ at $T=T_*$. On the fixed modulation length side of the crossover point, Re $k=q$.}
    \label{cixpoles}
  \end{center}
\end{figure*}

We will now consider the situation in which the 
system exhibits modulations at a fixed wave-vector $q$ 
for a finite range of temperatures on one side of $T_*$,
% [viz., {\bf (i)} $T_*\le T<T_*+\Delta T$, or, {\bf (ii)}  $T_*-\Delta T<T\le T_*$, $\Delta T>0$]
[viz., {\bf (i)} $T>T_*$, or, {\bf (ii)}  $T<T_*$]
and starts to exhibit variable wavelength modulations
on the other side [{\bf (iii)} $T<T_*$ for (i) and $T>T_*$ for (ii)].
A schematic illustrating this is shown in Fig. \ref{cixpoles}.
% The length scales from a real physical system can be seen in
% Fig. \ref{mesaros1b}.
In sub-section \ref{analxover}, we will assume that the pair correlation
function is meromorphic
(realized physically by {\em absence
  of phase transitions}) at the crossover point and illustrate how
modulation length exponents may appear.
In sub-section \ref{branchptxover}, we will comment on the situation
where the crossover point may be a branch point of the correlation function.

%%%%%%%%%%%%%%%%%%%%%%%%%%%%%%%%%%%%%%%%%%%%%%%%%%%%%%%%%%%%%%%%%%%%%%
\subsecn{Crossovers at general points in the complex
  $k$-plane}
%%%%%%%%%%%%%%%%%%%%%%%%%%%%%%%%%%%%%%%%%%%%%%%%%%%%%%%%%%%%%%%%%%%%%%
\label{analxover}
In the up and coming, we will assume that the pair correlator, $G(T,k)$ is a meromorphic
function of $k$ and $T$ near a crossover point.
Our analysis below is exact as long as we do not cross any phase boundaries.
Such a case is indeed materialized in the incommensurate-commensurate crossovers
in the three-dimensional
axial next-nearest-neighbor Ising
(ANNNI) model \cite{annni_first1, annni_first2} (which is of type (ii) in the classification above).
This phenomenon is also seen in the ground state phase diagram of Frenkel-Kontorova models \cite{fkmodels}
in which one of the coupling constants is tuned instead of
temperature.

In the following, we present two alternative derivations
for the  universal exponent characterizing this crossover.

%%%%%%%%%%%%%%%%%%%%%%%%%%%%%%%%%%%%%%%%%%%%%%%%%%
\subsubsecn{First approach}
%%%%%%%%%%%%%%%%%%%%%%%%%%%%%%%%%%%%%%%%%%%%%%%%%%
In general, if the pair correlation function $G(T,k)$
is a meromorphic function of the temperature $T$ and the wave-vector $k$ near a crossover
point $(T_*k_*)$, then $G^{-1}(T,k)$ must have a Taylor series expansion about
that point. We have,
\begin{eqnarray}
G^{-1}(T,k)= \sum_{m_1, m_2=0}^{\infty}A_{m_1m_2}(T-T_*)^{m_1}(k-k_*)^{m_2}.
\label{analyticGtk}
\end{eqnarray}
Since $G^{-1}(T_*,k_*)= 0$, we have, $A_{00}=0$.
Let us try to find the trajectory of the pole $K(T)$
(with $K(T_*)=k_*$) of
$G(T,k)$ in the complex $k$-plane as the temperature is varied
around $T_*$.
Writing down the leading terms of $G^{-1}(T,k)$ , we have, in general,
% \begin{eqnarray}
% G^{-1}(T,k)&\sim&
% A(T-T_*)^m(k-k_*)^n+\nonumber\\
% &&B(T-T_*)^{m-a}(k-k_*)^{n+b}+\nonumber\\
% &&\mbox{(More terms of the same order)}+\nonumber\\
% &&\mbox{(Smaller terms)},
% \end{eqnarray}
\begin{eqnarray}
G^{-1}(T,k)&\sim&\sum_{p=0}^{[m/a]}(T-T_*)^{m-ap}(k-k_*)^{n+bp}+\nonumber\\
&&~~~~~~~~o((T-T_*)^m(k-k_*)^n),
\end{eqnarray}
as $(T,k)\to(T_*,k_*)$ with $m,n,a,b$ integers, $m,n\ge0$ and
$a,b\ge1$, $[x]$ represents the greatest integer less than or equal to
$x$ and $o(x)$ represents terms negligible compared to $x$.
We have,
\begin{eqnarray}
% K(T)\sim k_*+\left(-A/B\right)^{1/b}(T-T_*)^{a/b},
K(T)\sim k_*+C(T-T_*)^{a/b},
\label{traj}
\end{eqnarray}
where $C$ is some constant,
yielding $\nu_L=a/b$.
By the very definition of $T_*$, on one side of $T_*$ [(i) or (ii) above],
there exists at least one root [say, $K(T)$] of $G^{-1}$ satisfying
$K_R(T)=q$, 
where $q$ is a constant.
On the other side [(iii) above], $K_R(T)\neq q$.
As such, the function $K(T)$ is non-analytic at $T_*$.
The left hand side of Eq. (\ref{traj}) is therefore not analytic at
$T=T_*$, implying that the right hand side cannot be analytic.
This means that $(a/b)$ cannot be an integer, which in turn implies that $b\ge2$.
Therefore, in the most common situations we might encounter, 
\begin{eqnarray}
&&G^{-1}(T,k)\sim A(T-T_*)+B(k-k_*)^2\nonumber\\
&\implies&a=1\mbox{ and }b=2.
\end{eqnarray}
When Fourier transforming $G(T,k)$ by evaluating the integral in
Eqs. (\ref{gkgxint}, \ref{gksphsym}) using the technique of residues, the real part
of the poles (i.e., $K_R$) gives rise to oscillatory modulations of
length $L_D=2\pi/K_R$.
If the modulation length locks its value to $2\pi/q$ on one side of the crossover point, 
then, on the other side, near $T_*$,
it must behave as
\begin{eqnarray}
|2\pi/L_D-q|\propto |T-T_*|^{1/2}\nonumber\\
\implies\boxed{\nu_L=1/2.}
\label{dle}
\end{eqnarray}

\begin{figure}
  \begin{center}
%    \vspace{0.2in}
%    \hspace{-.1in}
    \includegraphics[width=\columnwidth]{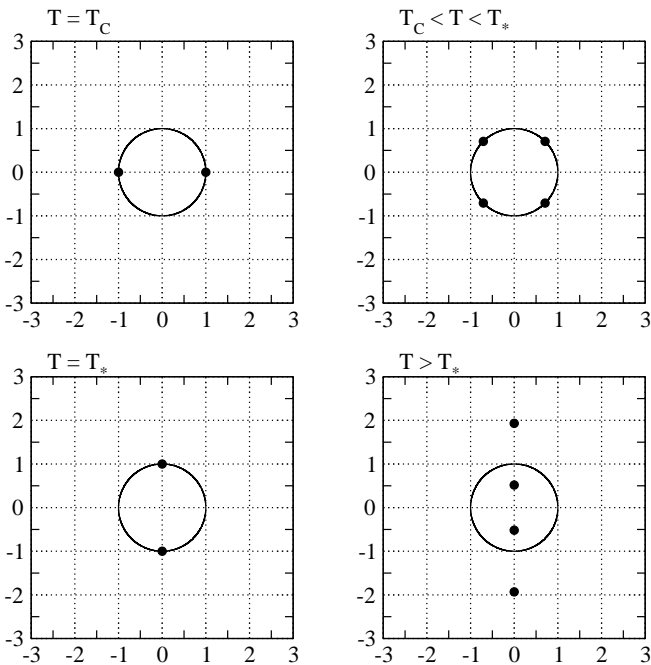}
    \caption{Location of the poles of the correlation function of the
      large $n$ Coulomb frustrated ferromagnet for $J=Q=1$
in the complex $k$-plane. The circle and the $Y$-axis show
the trajectory $K(T)$ of the poles as the temperature $T$ is varied.}
    \label{cfif_poles}
  \end{center}
\end{figure}

%%%%%%%%%%%%%%%%%%%%%%%%%%%%%%%%%%%%%%%%%%%%%%%%%%
\subsubsecn{Second approach}
%%%%%%%%%%%%%%%%%%%%%%%%%%%%%%%%%%%%%%%%%%%%%%%%%%
We now turn to a related alternative approach that similarly highlights the
universal character of the modulation length exponent.
If the correlation function $G$
is a meromorphic function of $k$, then,
expanding about a zero $K_1(T)$ of $G^{-1}$, we have,
\begin{eqnarray}
G^{-1}(T,k)= A(T) \left( k-K_1(T)\right)^{m_1}G_1^{-1}(T,k),
\label{analyticGtk0}
\end{eqnarray}
where $G_1^{-1}(T,k)$ is an analytic function of $k$ and $G_1^{-1}(T,K_1(T))\neq0$.
We can do this again for the function $G_1^{-1}(T,k)$ choosing one of
its zeros $K_2(T)$ and continue the process until the function left
over does not have any more zeros.
We have,
\begin{eqnarray}
% G^{-1}(T,k)= A(T) \prod_{a=1}^p\left( k-K_a(T)\right).
G^{-1}(T,k)= A(T) \prod_{a=1}^p\left( k-K_a(T)\right)^{m_a}G^{-1}_p(T,k),
\label{analyticGtk00}
\end{eqnarray}
where the function $G^{-1}_p(T,k)$ is an analytic function with no
zeros, $m_a$s are integers and, in principle, $p$ may be arbitrarily high.
This factorization can be done in each phase where $G$ is meromorphic.
Let $K_1(T)$ be a non-analytic zero of $G^{-1}$, i.e., one for
which $\mbox{Re }K_1(T)=q$ on one side of $T=T_*$.
To ensure analyticity of $G^{-1}$ in $T$ in the vicinity of $T=T_*$,  
there must be at least one other root $K_2(T)$, such that as $T\to T_*$, both $K_1(T)$ and $K_2(T)$
veer towards $k_*$, where ${\mbox Re}\ k_*=q$ [e.g., see
Fig. \ref{cfif_poles} which is of type (i) above, $k_*=\pm i$].
%\vspace{.5in}
In other words, $p$ in Eq. (\ref{analyticGtk00}) cannot be smaller than two.
The proof of this assertion is simple.
If $p=1$, then, according to Eq. (\ref{analyticGtk00}), $G^{-1}(T,k)=A(k-K_1(T))G_1^{-1}(T,k)$.
At $T=T_*$, however, $K_1(T)$ is not analytic, implying that $G^{-1}(T,k)$ can be analytic only if $p\ge2$.
For $p\ge 2$, at $T_*$, 
$G^{-1}$ will, to leading order, vary quadratically in $(k-k_*)$ in the complex $k$ plane
near $k_*$.
Thus,
\begin{eqnarray}
\left.\frac{\partial G^{-1}}{\partial k}\right|_{(T_*,k_*)}=0.
\end{eqnarray}
Now, if $G^{-1}$ has a finite 
first partial derivative relative to the temperature $T$ then, for a pole $K$ near $k_*$, to leading order,
\begin{eqnarray}
G^{-1}(T_*,k_*)+\left.(T-T_*)\frac{\partial G^{-1}}{\partial T}\right|_{(T_*,k_*)}& & \nonumber\\
+\left.\frac{(K-k_*)^2}{2!}\frac{\partial^2 G^{-1}}{\partial k^2}\right|_{(T_*,k_*)}&=&0.
\end{eqnarray}
By its definition, $k_*$ satisfies the equality
$G^{-1}(T_*, k_*) =0$. Therefore,
\begin{eqnarray}
|K-k_*|=\sqrt{\frac{2(T_*-T)\left.\frac{\partial G^{-1}}{\partial T}\right|_{(T_*,k_*)}}
{\left.\frac{\partial^2G^{-1}}{\partial k^2}\right|_{(T_*,k_*)}}}.
% k-k_*=\sqrt{2(T_*-T)\left(\left.\frac{\partial G^{-1}}{\partial T}\right|_{(T_*,k_*)}\right)/
% \left(\left.\frac{\partial^2G^{-1}}{\partial k^2}\right|_{(T_*,k_*)}\right)}
\end{eqnarray}

Equation (\ref{dle}) is an exact equality.
It demonstrates that the exponent $\nu_L=1/2$ {\em universally} unless one of $\frac{\partial^2 G^{-1}}{\partial k^2}$ 
and $\frac{\partial G^{-1}}{\partial T}$
vanishes at $(T_*,k_*)$.\cite{nussinov05} 
Often, $G^{-1}(T,k)$ is a rational function of $k$, i.e., 
\begin{eqnarray}
G^{-1}(T,k)=\frac{G^{-1}_n(T,k)}{G^{-1}_d(T,k)},
\end{eqnarray}
where $G^{-1}_n(T,k)$ and $G^{-1}_d(T,k)$ are polynomial functions of $k$.
In those instances, we get the same result as above by using
$G^{-1}_n(T,k)$ in the above arguments.
The value of the critical exponent is similar
to that appearing for the correlation length exponent for mean-field or
large $n$ theories. It should be stressed that our result of Eq. (\ref{dle})
is far more general.

{\em Lock-in of the correlation length.}
Apart from the crossovers across which the modulation length locks in
to a fixed value, we can also have situations where
the correlation length becomes constant as a crossover temperature $T_{**}$ is crossed.
If this happens, our earlier analysis for the modulation length may be
replicated anew for the correlation length.
Therefore, if the correlation length has a fixed value $\xi_0$ on one
side ($T<T_{**}$ or $T>T_{**}$) of the crossover point, 
then, on the other side ($T>T_{**}$ or $T<T_{**}$, respectively), near $T_{**}$,
it must behave as,
\begin{eqnarray}
|1/\xi-1/\xi_0|\propto|T-T_{**}|^{\nu_c},
\label{xitst}
\end{eqnarray}
where, like $\nu_L$, $\nu_c=1/2$ apart from special situations where
it may take some other rational values.
Here and throughout, we use $\nu_c$ to represent the usual correlation length
exponent, $\nu$ to distinguish it from the modulation length exponent $\nu_L$.

%%%%%%%%%%%%%%%%%%%%%%%%%%%%%%%%%%%%%%%%%%%%%%%%%%%%%%%%%%%%
\subsecn{Branch points}\label{branchptxover}
%%%%%%%%%%%%%%%%%%%%%%%%%%%%%%%%%%%%%%%%%%%%%%%%%%%%%%%%%%%%
A general treatment of a situation in which the crossover point is a branch point of the inverse correlation function in the
complex $k$-plane is beyond the scope of this work. 
Branch points are ubiquitous in correlation functions in both
classical as well as quantum systems.

For example, in the large $n$ rendition of a bosonic system
(with a Hamiltonian of Eq. (\ref{Ham}) and $S(x)$ representing bosonic
fields), the
momentum space correlation function at temperature $T$ is given by
\cite{ref:nussinovAPT, largen} 
\begin{eqnarray}
G(\vec{k})=\sqrt{\frac{\mu_1}{v(\vec{k})+\mu}} \left[ n_B\left( \frac{\sqrt{\mu_1(v(\vec{k})+\mu)}}{k_BT} \right) + \frac{1}{2}\right],
\end{eqnarray}
where $\mu_1$ is a constant having dimensions of energy, 
$\mu$ is the chemical potential, 
$n_B(x) = 1/(e^x-1)$ is the Bose distribution function and 
$k_B$ is the Boltzmann's constant.

Similar forms, also including spatial modulations in $G(r)$, may also appear.
We briefly discuss examples where we have a branch cut in the complex
$k$-plane.

The one-dimensional momentum space correlation function,
\begin{eqnarray}
G(k)=\frac{1}{\sqrt{(k-q)^2+r}}+\frac{1}{\sqrt{(k+q)^2+r}},
\end{eqnarray}
reflects a real space correlation function given by
\begin{eqnarray}
G(x)=\frac{2\cos(qx)\mbox{K}_0(x\sqrt{r})}{\pi},
\end{eqnarray}
where $\mbox{K}_0(\cdot)$ is a modified Bessel function. Thus, as is to be expected, we obtain length scales
associated with the branch points $K=\pm q\pm i\sqrt{r}$.

Similarly, the three-dimensional real space correlation function corresponding to
\begin{eqnarray}
G(k)=\frac{1}{\sqrt{(k-q)^2+r}},\label{sqrtgk}
\end{eqnarray}
exhibits the same correlation and modulation lengths along with an
algebraically decaying term for large separations.
Another related $G^{-1}(k)$ involving a function of $|\vec{k}|$
(i.e., not an analytic  function of $k^{2}$) was investigated earlier.\cite{c29avoided}

Throughout the bulk of our work, we consider simple exponents associated with analytic crossovers. In considering brach points, our analysis may be extended to critical points. As is well known, at critical points of $d$ dimensional systems, the correlation function for large $r$, scales as
\begin{eqnarray}
G(r)\propto\frac{1}{r^{d-2+\eta}},
\label{anomr}
\end{eqnarray}
with $\eta$ the {\em anomalous} exponent.
Such a scaling implies, for non-integer $\eta$, the existence of a $k=0$ branch point of $G(k)$. 

If the leading order behavior of $1/G^{(m)}(T,k)$ is algebraic
near a branch point $(T_*,k_*)$, then we get an algebraic exponent
characterizing a crossover at this point [$m$ being the lowest order derivative
of $G(k)$ which diverges at $k=k_*$ as in Eq. (\ref{lengthsK})]. That is, we have,
\begin{eqnarray}
\frac{1}{G^{(m)}(T,k)}&\sim& A(T-T_*)^{z_1}-B(k-k_*)^{z_2}\nonumber\\
&&\mbox{ as }(T,k)\to(T_*,k_*).\label{alphagamma}
\end{eqnarray}
This implies that the branch points $K$ deviate from $k_*$ as
\begin{eqnarray}
(K-k_*)\sim\left(\frac{A}{B}\right)^{1/{z_2}}(T-T_*)^{{z_1}/{z_2}}.
\end{eqnarray}
We therefore observe a length scale exponent $\nu={z_1}/{z_2}$ at this crossover.
This exponent may characterize a correlation lengths and/or a modulation lengths.
The exponent $z_1/z_2$ may assume {\em irrational} values in many
situations in which the function $G^{-1}(T,k)$ is not analytic near
the crossover point.
Such a situation could give rise to phenomena exhibiting anomalous
exponents $\eta$.
For example, if we have a diverging correlation length at a critical temperature
$T_c$, for a system with a correlation function which behaves as in
Eq. (\ref{anomr}), then, we have in Eq. (\ref{alphagamma}),
${z_2}=2-\eta$.
Thus, we have,
\begin{eqnarray}
|L_D-L_{Dc}|\propto|T-T_c|^{\frac{z_1}{2-\eta}},\nonumber\\
\implies \boxed{\nu_L=\frac{z_1}{2-\eta}},
\label{nulanom}
\end{eqnarray}
where $L_{Dc}=2\pi/|\mbox{Re }k_*|$, 
and more importantly,
\begin{eqnarray}
\xi\propto|T-T_c|^{-\frac{z_1}{2-\eta}},\nonumber\\
\implies \boxed{\nu_c=\frac{z_1}{2-\eta}}.
\label{nucanom}
\end{eqnarray}
Other critical exponents could also, in principle, be calculated using hyper-scaling relations.

If $G^{-1}(T,k)$ % in Eq. (\ref{lengthsK}) 
has a Puiseux representation about the crossover point, i.e.,
\begin{eqnarray}
G^{-1}(T,k)=\sum_{m=m_0}^{\infty}\sum_{p=p_0}^{\infty}a_{mp}(k-k_*)^{m/a}(T-T_*)^{p/b},
\end{eqnarray}
with $a_{m_0p_0}=0$, where $m_0,~p_0,~a$ and $b$ are integers,
then, the result we derived above applies to the relevant length scale and
the crossover exponent $\nu=a/b$, is again a {\em rational number}.

Generalizing, if $G^{-1}(T,k)$ is the ratio of two Puiseux series, we
use the numerator to obtain the leading order asymptotic behavior and
hence obtain a rational exponent.

%%%%%%%%%%%%%%%%%%%%%%%%%%%%%%%%%%%%%%%%%%%%%%%%%%%%%%%%%%%%
\subsecn{A corollary: Discontinuity in modulation lengths implies
  a thermodynamic phase transition} 
%%%%%%%%%%%%%%%%%%%%%%%%%%%%%%%%%%%%%%%%%%%%%%%%%%%%%%%%%%%%
% As the wave-vectors characterizing a system are continuous at
% crossover points near which the inverse
% correlation function in Fourier space is analytic in $T$ and $k$,
% situations where there are discontinuous ({\em first order}) jumps in the
% correlation or modulation lengths must correspond to non-analyticities in the inverse correlation function suggesting the
% presence of a phase transition. 
Non-analyticities in the correlation function $G(k)$ for real
wave-vector $k$ imply the existence of a phase transition. This leads
to simple corollaries as we now briefly elaborate on. A sharp
discontinuous jump in the value of the modulation lengths (and/or
correlation lengths) implies that the zeros $\{K_{a}\}$ of $G^{-1}(k)$
in the complex $k$ plane, exhibit discontinuous (``first order-like'')
jumps as a function of some parameter (such as the temperature $T$
when $T=T_*$). When this occurs, as seen by, e.g., differentiating the
reciprocal of the product of Eq. (\ref{analyticGtk00}), the correlation function
will, generally, not be analytic as a function of $T$ at
$T=T^*$. Putting all of the pieces together, we see that a
discontinuous change in the modulation (or correlation) lengths impies
the existence of a bona fide phase transition. Thus, all
commensurate-commensurate crossovers must correspond to phase transitions.
For example, see the ANNNI model.\cite{annni_rev}

%%%%%%%%%%%%%%%%%%%%%%%%%%%%%%%%%%%%%%%%%%%%%%%%%%
\subsecn{Diverging correlation length at a spinodal transition}
%%%%%%%%%%%%%%%%%%%%%%%%%%%%%%%%%%%%%%%%%%%%%%%%%%
Our %$\vec{k}$-space 
analysis is valid for both annealed and quenched
systems so long as translational symmetry is maintained
(and thus, the correlation function is diagonal in $k$-space). In
particular, whenever phase transitions are ``avoided'' the rational
exponents of Eq. (\ref{traj}) will appear.\cite{zohar, us, ref:nussinovAPT}

In diverse arenas, we may come across situations in which there are
no diverging correlation lengths even when the inverse correlation
function has zeros corresponding to real values of the
wave-vector. These are signatures of a first order phase transition,
e.g., transition from a liquid to a crystal.
If the first order phase transition is somehow avoided, then the system
may  enter a metastable phase and may further reach a point where the
correlation length diverges, e.g., a spinodal point. If it is possible to reach this
point and if the inverse correlation function is analytic there, then our
analysis will be valid, thereby leading to rational exponents
characterizing the divergence of the correlation
length. There are existing works in the literature which seem
to suggest that such a point may not be reachable. For example, in
mode coupling theories of the glass transition, the system reaches
the mode coupling transition temperature $T_{MCT}$ at which the
viscosity and relaxation times diverge and hence does not reach the
point where the correlation length
blows up.\cite{kirkthirdynglass}

%\textcolor{red}{
\subsecn{Conservation of characteristic length scales}\label{corrmodcons}
In Ref. \cite{largen}, it was mentioned that the total number of
characteristic length scales in a large-$n$ system remains constant in
systems in which the Fourier space interaction kernel $v(\vec{k})$ is a
rational function of $k^2$ and the real space kernel is rotationally
invariant. (Similar results hold for systems with reflection point group symmetry.\cite{transrotinvariance})
In this sub-section, we generalize that argument and say that whenever the
Fourier space correlator $G(\vec{k})$ of a general rotationally invariant system is a rational function
of $k^2$,i.e.,
\begin{eqnarray}
G(\vec{k})=\frac{P(k^2)}{Q(k^2)},
\label{gkratpq}
\end{eqnarray}
the total number of correlation and modulation lengths
remains constant apart from isolated points as a tuning parameter
$\lambda$ is smoothly varied. In Eq. \ref{gkratpq}, the functions $P(k^2)$
and $Q(k^2)$ are polynomial functions of $k^2$. 
Rotational invariance requires that $G(\vec{k})$ is real-valued for real
wavevectors $k$.
As argued in Ref. \cite{largen}, all length scales in
the such systems are associated with the poles of $G(k)$ in the complex
$k$-plane and these remain constant for a given form of the function
$G(k)$. Each real root of the function $Q(k^2)$ gives rise to a term
in the real space correlation function which has one
correlation or modulation length. Non-real roots (which necessarily
come in complex conjugate pairs) give rise to a correlation and a
modulation length. Thus, the total number of characteristic length
scales in the system is equal to the order of the polynomial function
$Q(k^2)$ which remains fixed.
%}%color

%%%%%%%%%%%%%%%%%%%%%%%%%%%%%%%%%%%%%%%%%%%%%%%%%%%%%%%%%%%%%%%%%%%%%%%%%%%%%%%%
\secn{$O(n)$ systems}
%%%%%%%%%%%%%%%%%%%%%%%%%%%%%%%%%%%%%%%%%%%%%%%%%%%%%%%%%%%%%%%%%%%%%%%%%%%%%%%%
The correlation function for $O(n)$ systems can be calculated exactly
at both the low and the high temperature limits. At intermediate temperatures,
various crossovers and phase transitions may appear.
In this section, we discuss the low and high temperature behavior
length scales characterizing $O(n)$ systems.

%%%%%%%%%%%%%%%%%%%%%%%%%%%%%%%%%%%%%%%%%%%%%%%%%%%%%%%%%%%%
\subsecn{Low temperature configurations}\label{ltc} 
%%%%%%%%%%%%%%%%%%%%%%%%%%%%%%%%%%%%%%%%%%%%%%%%%%%%%%%%%%%%
It  was earlier demonstrated \cite{zohar_com} that for $O(n\ge2)$, {\em all} ground states of a system have to be 
spirals (or poly-spirals) of characteristic wave-vectors
$\vec{q}_{\alpha}$, given by
\begin{eqnarray}
v(\vec{q}_{\alpha})=-\min_{\vec{k}\in\mathbb{R}^d}v(\vec{k}),
\end{eqnarray}
where $\mathbb{R}^d$ represents the set of all $d$-dimensional real vectors.
At $T=0$, the modulation lengths in the system are given by
\begin{eqnarray}
L_D^{i,\alpha}(T=0)=2\pi/q_{i,\alpha},
\end{eqnarray}
where $i$($1\le i\le d$) labels the Cartesian directions in $d$ dimensions.
This, together with Eq. (\ref{ht}) gives us the high and low temperature forms of the correlation function and its
associated length scales.

%%%%%%%%%%%%%%%%%%%%%%%%%%%%%%%%%%%%%%%%%%%%%%%%%%%%%%%%%%%%
\subsecn{High temperatures}\label{secht}
%%%%%%%%%%%%%%%%%%%%%%%%%%%%%%%%%%%%%%%%%%%%%%%%%%%%%%%%%%%%
As is well appreciated, diverse systems behave in the same way at high temperatures.\cite{hightemp}
For $O(n)$ systems \cite{onsystem} (any $n$),
\begin{eqnarray}
G^{-1}(T,k)=1+v(k)/k_BT+{\cal O}(1/T^3).
\label{ht}
\end{eqnarray}
The high temperature series may be extended and applied at the crossover temperature $T_*$,
if there is no phase transition at temperatures above $T_*$ and for all relevant real
$k$'s, $|v(k)|\ll k_BT_*$. [A detailed example will be studied in Sec. \ref{exhightemp}.]
Generally, Eq. (\ref{ht}) may be analytically continued for complex $k$'s and in the vicinity of $T_*$,
\begin{eqnarray}
\delta k \sim \left[\frac{m!~k_B(T_*-T)}{v^{(m)}(k_*)}\right]^{\frac{1}{m}},
\end{eqnarray}
where $k_*$ is a characteristic wave-vector at $T_*$. In the above, $\delta k$ denotes the change in the location of the poles $K$ of $G^{-1}$ when 
the temperature is changed from $T_*$ to $T$ (i.e., $\delta k\equiv K-k_*$)
and $m$ is the order of the lowest non-vanishing derivative of $v(k)$ at $k_*$.
As in previous analysis, $v'(k_*)=0$ and $m\ge2$.
For general $v(k)$, typically $m=2$ and $\nu_L=1/2$ as before.

We now turn to examples which explicitly illustrate how our results are realized including exceptional systems with non-trivial exponents.

%%%%%%%%%%%%%%%%%%%%%%%%%%%%%%%%%%%%%%%%%%%%%%%%%%%%%%%%%%%%
\subsecn{Large $n$ Coulomb frustrated ferromagnet -- modulation length
exponent at the crossover temperature $T_*$}
%%%%%%%%%%%%%%%%%%%%%%%%%%%%%%%%%%%%%%%%%%%%%%%%%%%%%%%%%%%%
In the current sub-section and the two that follow, we will discuss the large $n$ limit
in $O(n)$ systems. The results in the previous two sections pertain to
arbitrary $n$.
We illustrate how our result applies to the large $n$ \cite{onsystem} Coulomb frustrated ferromagnet.
As is well known \cite{stanley}, in the large $n$ limit, $O(n)$
systems are exactly solvable and behaves as the spherical model.\cite{kac}
The correlation function in $k$-space is given by
\begin{eqnarray}
G^{-1}(T,k)=[v(k)+\mu(T)]/k_BT,
\label{gmt}
\end{eqnarray}
where $v(k)$ is the Fourier space interaction kernel and
$\mu(T)$ is a Lagrange multiplier, see e.g. Ref. \cite{largen, us}, that enforces the spherical
constraint
\begin{eqnarray}
\frac{1}{N}\sum_{\vec{x}}\langle\vec{S}(\vec{x})\cdot\vec{S}(\vec{x})\rangle=1.
\end{eqnarray}
The paramagnetic transition temperature $T_C$ is obtained from the relation,
$\mu(T_C)=-\min_{k\in\mathbb{R}}v(k)$.
Below $T_C$, the Lagrange multiplier $\mu(T)=\mu(T_C)$.
Above $T_C$, $\mu(T)$ is determined by the global average constraint that
$G(\vec{x}=0)=\frac{1}{N}\sum_{\vec{k}}G(\vec{k})=1$.
This global constraint also implies that, above $T_C$, 
small changes in temperature result in proportional changes in $\mu(T)$
and at high temperatures,
$\mu(T)$ is a monotonic increasing function of $T$.
The Fourier space kernel $v(k)$ for the ``Coulomb frustrated
ferromagnet'' (in which nearest neighbor ferromagnetic interactions of
strength $J$ compete with Coulomb effects of strength $Q$) is given by
$v(k)=Jk^2+Q/k^2$, where $J$ and $Q$ are positive constants.
The critical temperature, $T_C$ of this system is given by
$ %\begin{eqnarray}
\mu(T_C)=-2\sqrt{JQ}.
$ %\end{eqnarray}
At $T_C$, the correlation length is infinity and the modulation length is $L_D=2\pi\sqrt[4]{J/Q}$.
As the temperature is increased, the modulation length increases and the correlation length decreases.
At $T_*$, given by $\mu(T_*)=2\sqrt{JQ}$, the modulation length diverges and the correlation length becomes
$\xi=\sqrt[4]{J/Q}$.
At temperatures above $T_*$, the correlation function exhibits no modulations and there is one decreasing correlation length
and one increasing correlation length. The term in the correlation function with the increasing correlation length
becomes irrelevant at high temperatures because of an algebraically decaying prefactor.
The divergence of the modulation length at $T_*$ shows an exponent of $\nu_L=1/2$.\cite{largen}

\begin{figure}
  \begin{center}
    \vspace{.2in}
%    \hspace{-1in}
    \includegraphics[width=\columnwidth]{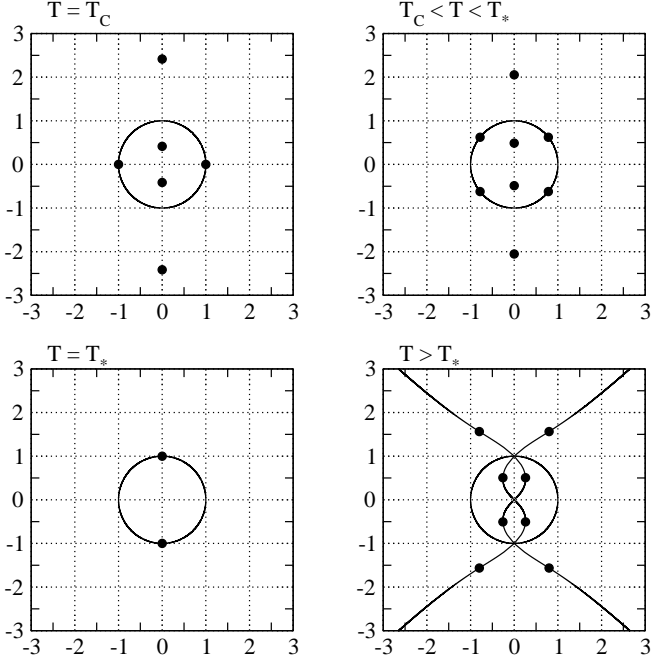}
    \caption{Location of the poles of the correlation function of the system in Eq. (\ref{exv}) for large $l_s$ (small screening) 
in the complex $k$-plane.}
    \label{example_poles}
  \end{center}
\end{figure}

%%%%%%%%%%%%%%%%%%%%%%%%%%%%%%%%%%%%%%%%%%%%%%%%%%%%%%%%%%%%
\subsecn{An example with $\mathbf{\nu_L\neq1/2}$}\label{nu.25}
%%%%%%%%%%%%%%%%%%%%%%%%%%%%%%%%%%%%%%%%%%%%%%%%%%%%%%%%%%%%
In what follows, we demonstrate, as a matter of principle, that the exponent for the divergence of the modulation length
(and also the correlation length) can be different from $1/2$ in certain special cases.
As an illustrative example, we consider a large $n$ (or spherical model) system for which
in Eq. (\ref{Hamkspace}),
\begin{eqnarray}
%v(k)=k^4+4k^2+4k^{-2}+k^{-4}.
v(k)&=&A(k^2+l_s^{-2})^2+4B(k^2+l_s^{-2})\nonumber\\
&&+4C/(k^2+l_s^{-2})+D/(k^2+l_s^{-2})^2,
\label{exv}
\end{eqnarray}
where $l_s$ is a screening length.
If we set $A=B=C=D=1$ then in the resultant system $\nu_L\neq1/2$ at a crossover temperature.
It has a critical temperature $T_C$, given by
$ %\begin{eqnarray}
\mu(T_C)=-10.
$ %\end{eqnarray}
At $T_C$, the modulation length is $L_D=2\pi/\sqrt{1-1/l_s^2}$ and the correlation length blows up (as required by definition).
At the crossover temperature $T_*$ (for which
$\mu(T_*)=6$) %\end{eqnarray}, 
the modulation length diverges and the correlation length scales as $\xi=1/\sqrt{1+1/l_s^2}$.
A temperatures just below $T_*$, the modulation length $L_D$ diverges as $L_D\propto(T_*-T)^{-1/4}$ meaning that $\nu_L=1/4$.
This is because the first three derivatives of $v(k)$ vanish at $k=i$, which is the characteristic wave-vector at $T_*$
(see Fig. \ref{example_poles}).
% \begin{figure}[h]
%   \begin{center}
%     \vspace{0.2in}
%     \includegraphics[width=\columnwidth]{example_poles}
% %    \epsfig{file=example_poles,width=\columnwidth}
%     \caption{Location of the poles of the correlation function of the system in Eq. (\ref{exv}) for large $l_s$ (small screening) 
% in the complex $k$-plane.
% At $T_C$ the poles are at the points labeled by A. A1 and A2 are poles of second order and the rest are simple poles.
% As the temperature is increased, each of the second order poles split into two simple poles which move along the unit circle towards the imaginary axis. The poles which are on the imaginary axis stay imaginary and move towards the unit circle.
% At $T_*$, the poles reach the B-points given by $k=\pm i$ and the poles are of fourth order.
% As the screening is increased, the circle distorts so that A1 and A2 move closer to the origin and B1 and B2 move away.}
%     \label{example_poles}
%   \end{center}
% \end{figure}

% \begin{figure*}[t]
%   \begin{center}
%     \vspace{0.2in}
%     \includegraphics[width=\textwidth]{espoles}
%     \caption{Location of the poles of the correlation function of the system in Eq. (\ref{exv}) for large $l_s$ (small screening) 
% in the complex $k$-plane.}
%     \label{example_poles}
%   \end{center}
% \end{figure*}

%%%%%%%%%%%%%%%%%%%%%%%%%%%%%%%%%%%%%%%%%%%%%%%%%%%%%%%%%%%%
\subsecn{An example in which $T_*$ is a high temperature}\label{exhightemp}
%%%%%%%%%%%%%%%%%%%%%%%%%%%%%%%%%%%%%%%%%%%%%%%%%%%%%%%%%%%%
We now provide an example in which the high
temperature result of Sec. \ref{secht} (valid for any $O(n)$ system with arbitrary $n$) can be applied at a crossover point.
Consider the large $n$ system in Eq. (\ref{exv}) with $A=1$, $B\gg1$, $C=1/4$, $D=0$ and the screening length,
$l_s\gg B$.
The critical temperature of this system is given by
$
\mu(T_C)\sim-4\sqrt{B}
$
where the modulation length is $L_D\sim2\pi\sqrt[4]{4B}$.
There is a crossover temperature $T_*$ such that
$
\mu(T_*)\sim 4B^2.
$
One of the modulation lengths diverges at $T_*$.
The corresponding correlation length is given by $\xi\sim1/\sqrt{2B}$.
This provides
an example in which $|v(k)|\ll k_BT_*$ for all real $k$'s satisfying $|k|\le\pi$.
The second derivative of $v(k)$ is non-zero at the crossover point, resulting in a crossover exponent $\nu_L=1/2$.

\begin{figure}
  \begin{center}
    \vspace{.2in}
%    \hspace{-1in}
\includegraphics[width=.45\columnwidth]{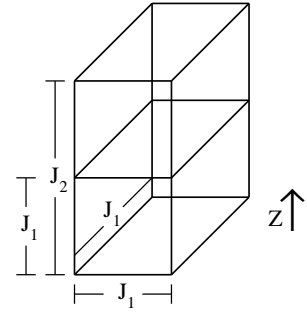}
    \caption{
The coupling constants in the three-dimensional ANNNI model.
}\label{annnicoup}
  \end{center}
\end{figure}

\begin{figure}
  \begin{center}
%    \vspace{.2in}
%    \hspace{-1in}
%\subfigure[]{\includegraphics[width=.45\columnwidth]{annnij1j2.eps}\label{annnicoup}}\\
\subfigure[]{\includegraphics[width=.45\columnwidth]{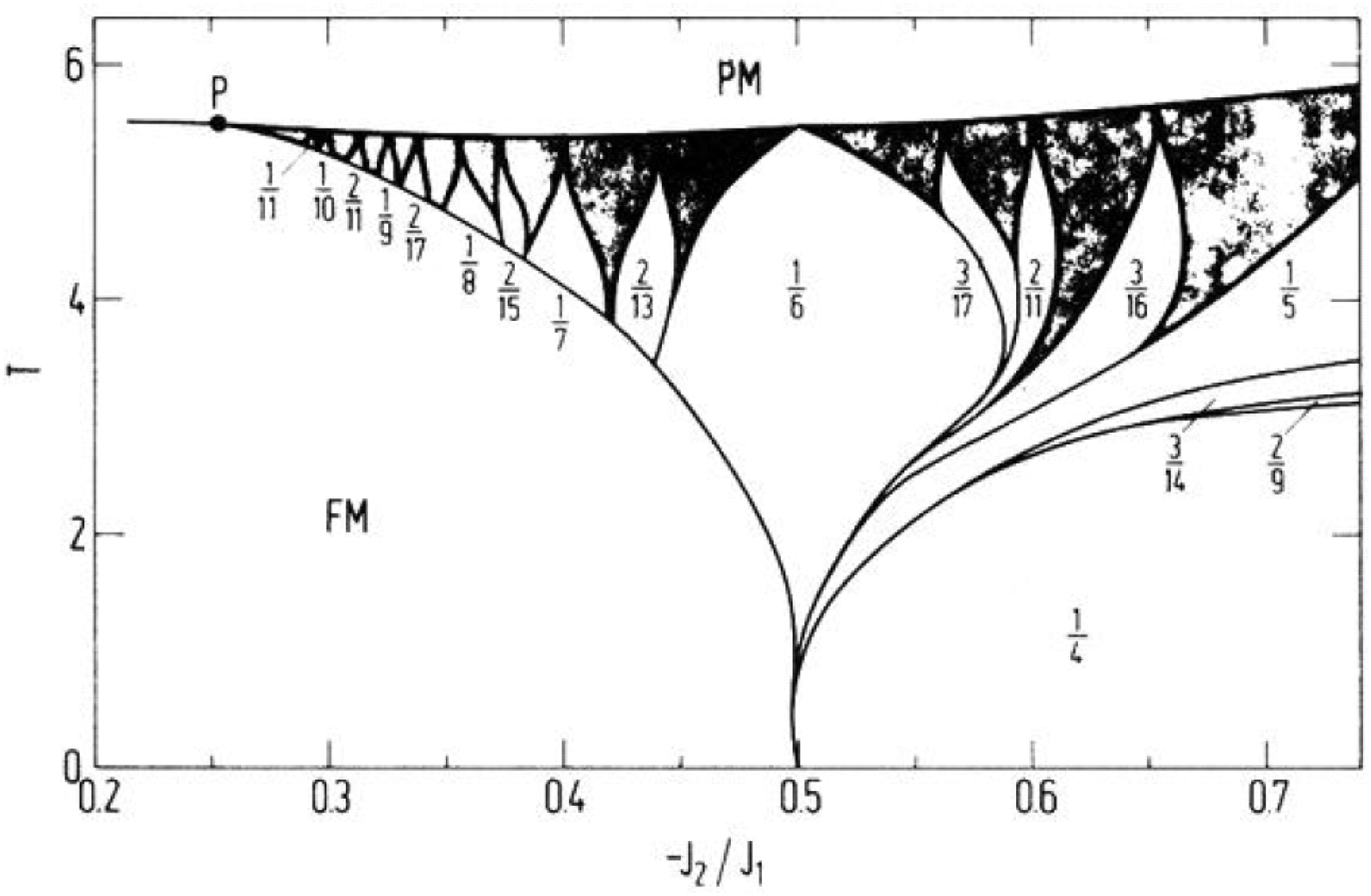}\label{bb13}}\qquad
\subfigure[]{\includegraphics[width=.45\columnwidth]{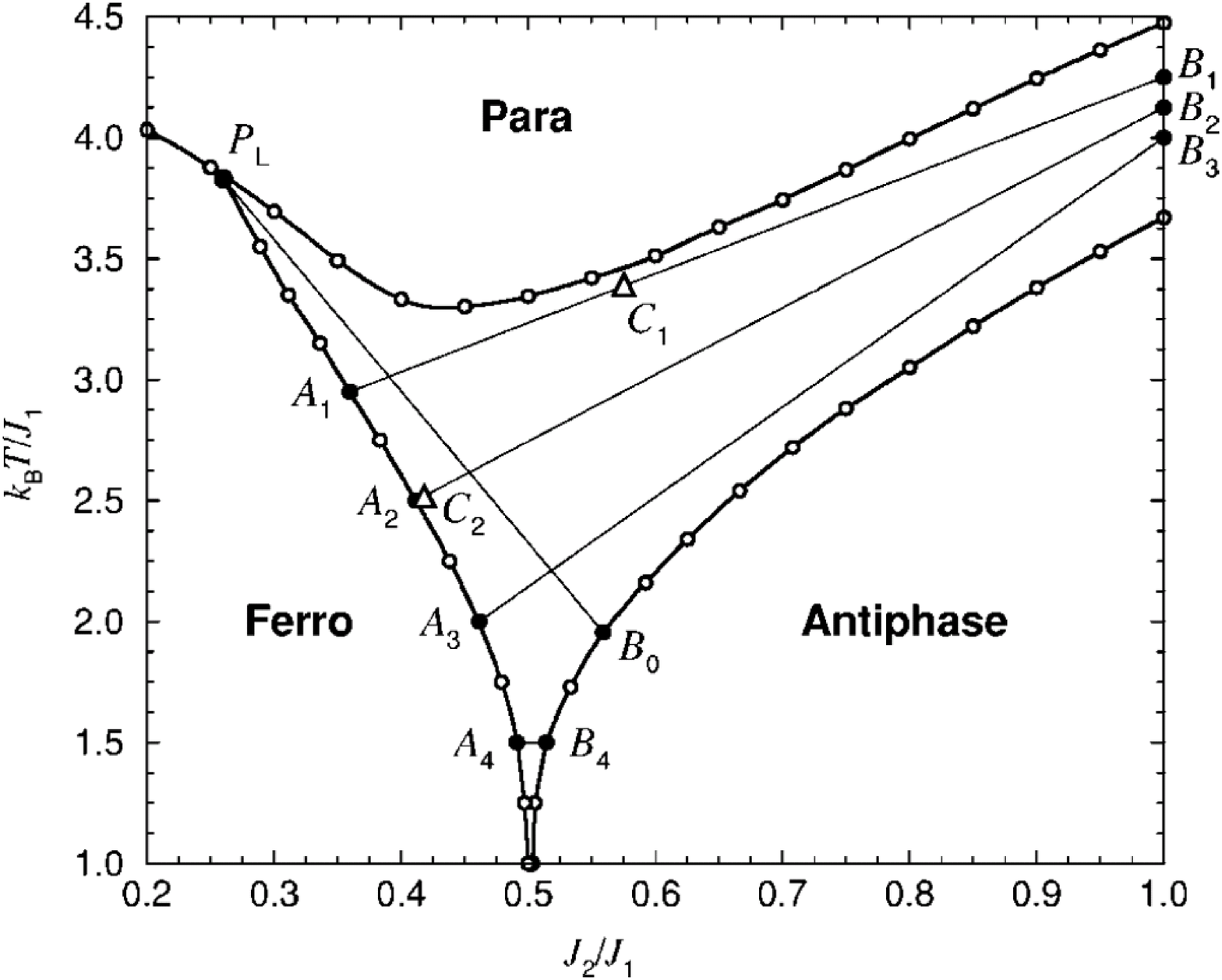}\label{g4}}\\
\subfigure[]{\includegraphics[width=.45\columnwidth]{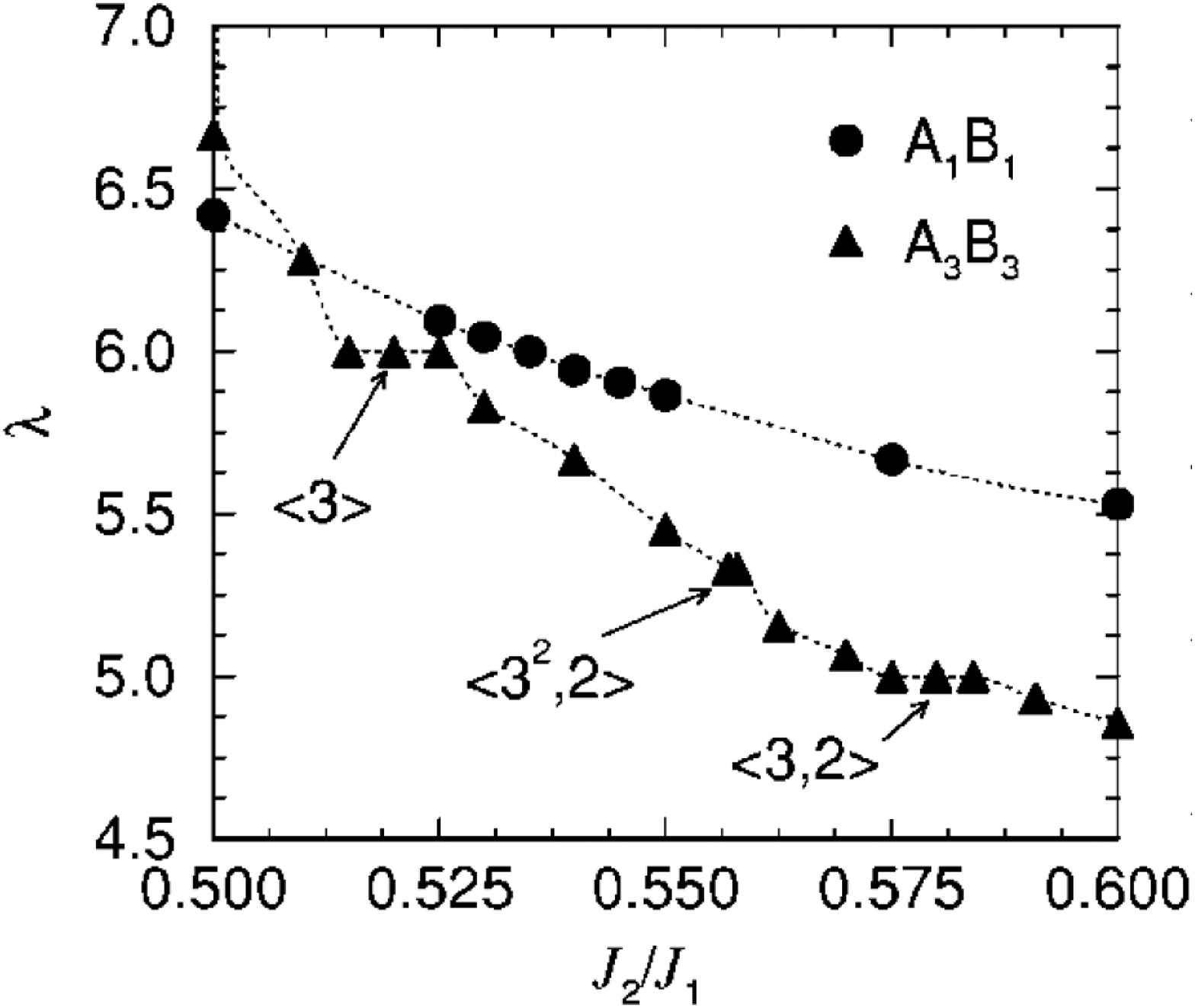}\label{g7}}\qquad
\subfigure[]{\includegraphics[width=.45\columnwidth]{iccartoon.eps}\label{icc}}
    \caption{
Existence of incommensurate phases between the commensurate regions in
the phase diagram of the ANNNI model.
{\bf \subref{bb13}} Mean field phase diagram of the ANNNI model in three dimensions. The shaded regions show higher order 
commensurate phases which have variable modulation length incommensurate phases in between  (From
Ref. \cite{bakboehm}. Reprinted with permission from APS.)
{\bf \subref{g4}} Phase diagram for the three-dimensional ANNNI model using a modified tensor product variational approach (From Ref. \cite{gendiar}. Reprinted with permission from APS.)
{\bf \subref{g7}} Variation of wavelength
along paths $A_1B_1$ and $A_3B_3$ of \subref{g4} showing a smooth
variation of the wavelength near the paramagnetic transition line (From Ref. \cite{gendiar}. Reprinted with permission from APS.)
{\bf \subref{icc}} Cartoon of an incommensurate-commensurate crossover region from \subref{bb13}.
}
    \label{annni_old}
  \end{center}
\end{figure}

%%%%%%%%%%%%%%%%%%%%%%%%%%%%%%%%%%%%%%%%%%%%%%%%%%%%%%%%%%%%
\secn{Crossovers in the ANNNI model}
%%%%%%%%%%%%%%%%%%%%%%%%%%%%%%%%%%%%%%%%%%%%%%%%%%%%%%%%%%%%
We now comment on one of the oldest studied examples of a system with a crossover temperature.
The following Hamiltonian represents the ANNNI model.\cite{annni_first1, annni_first2, annni_rev}
\begin{eqnarray}
H=-J_1\sum_{\langle \vec{x},\vec{y}\rangle}S(\vec{x})S(\vec{y})+J_2\sum_{\langle\langle \vec{x},\vec{y}\rangle\rangle}S(\vec{x})S(\vec{y}),
\end{eqnarray}
where 
as throughout, $\vec{x}$ is a lattice site on a cubic lattice,
and the spins $S(\vec{x})=\pm1$.
The couplings, $J_1,J_2>0$.
In the summand, $\langle\cdot\rangle$ represents nearest neighbors and $\langle\langle\cdot\rangle\rangle$ represents
next nearest neighbors along one axis (say the $Z$-axis), see Fig. \ref{annnicoup}.
Depending on the relative strengths of $J_1$ and $J_2$,
the ground state may be either ferromagnetic or in the
``$\langle2\rangle$ phase''. The ``$\langle2\rangle$ phase'' is a
periodic layered phase, in which there are layers of width two lattice
constants of `up'' spins alternating with layers of ``down'' spins of
the same width, along the $Z$-axis.
% that alternates two ``layers'' of pluses with two layers of minuses along the $Z$-axis.
As the temperature is increased, the correlation function exhibits jumps in the modulation wave-vector
at different temperatures.
At these temperatures, the system undergoes first order transitions to different commensurate phases.
The inverse correlation function $G^{-1}(T,k)$ is therefore not an analytic
function of $k$ and $T$ at the transition points.
The phase diagram
for the ANNNI model, however, also has several crossovers where the system
goes from a commensurate phase to an incommensurate phase with a
continuously varying modulation length (see Fig. \ref{annni_old}).\cite{bakboehm,gendiar}
At these crossover points, following our rigorous analysis, we expect a crossover exponent $\nu_L=1/2$.
Such a scaling of the modulation length has been estimated by several approximate techniques near the ``Lifshitz point'' $P_L$.\cite{annni_rednerstanleyjphysc, annni_rednerstanleyprb, annni_oitmaa,
  annni_mukamel, annni_hornreich, annni_rev,charbonneauprb}
The Lifshitz point is the point in the phase diagram of the ANNNI
model at which the high temperature paramagnetic phase coexists with
the ferromagnetic phase as well as a phase with continuously varying
modulation lengths. It  is marked as $P_L$ in Fig. \ref{g4}.
Although the point $P_L$ has a first order transition, it can be thought of as a limit in which the incommensurate and commensurate regions
in Fig. \ref{bb13} shrink and merge to a single point.
We would also like to point out that it is known \cite{bakreview}
that if the wave-vector takes all possible rational values (``complete devil's staircase''),
we have no first order transitions.
Additionally, non-analyticity of the correlation function does not prohibit other quantities from having continuous
crossover behavior. For example, the correlation of the fluctuations, i.e., the connected correlation function may
generally exhibit continuous variation from a fixed to a variable
modulation length phase.
If the inverse connected correlation function is analytic, our result can be applied to it resulting in a crossover exponent
of $1/2$.

% In summary, we find a new divergence 
% of the modulation length at the ``disorder'' line temperature
% $T_{disorder}$ or $T_*$ .
% \cite{nussinov05} 
% The value of the critical exponent is similar
% to that appearing for the correlation length exponent for mean-field or
% large $n$ theories. It should be stressed that our result of Eq. (\ref{dle})
% is far more general.

%%%%%%%%%%%%%%%%%%%%%%%%%%%%%%%%%%%%%%%%%%%%%%%%%%%%%%%%%%%%
\secn{Parameter extensions and generalizations} 
%%%%%%%%%%%%%%%%%%%%%%%%%%%%%%%%%%%%%%%%%%%%%%%%%%%%%%%%%%%%
It is illuminating to consider simple generalizations of our result to other arenas.
We may also replicate the above derivation for a system in which,
instead of temperature,
some applied other field $\lambda$ is responsible for the changes in the 
correlation function of the system. Some examples could be pressure, applied
magnetic field and so on. The complex wave-vector $k$ could also be
replaced by a frequency $\omega$ whose imaginary
part would then correspond to some decay constant in the time domain. 

More generally, we look for solutions to the equation
\begin{eqnarray}
G^{-1}(\lambda,u)=0,
\label{Gu}\end{eqnarray}
with the variable $u$ being a variable Cartesian component of the wave-vector, 
the frequency, or any other momentum space coordinate appearing in the
correlation function between two fields ($u= k_{i}, \omega$, and so on).
Replicating our steps mutatis mutandis, we find that
the zeros of Eq. (\ref{Gu}) scale as
$ %\begin{eqnarray}
|u-u_0|\propto|\lambda-\lambda_*|^{1/2}
$ %\end{eqnarray}
whenever the real (or imaginary) part of some root becomes constant as $\lambda$ crosses $\lambda_*$.
Thus, our predicted exponent of $\nu_L=1/2$ could be observed in a vast variety of systems
in which a crossover occurs as the applied field crosses a particular
value, in the complex wave-vector like variable.

Another generalization of our result proceeds as follows.\cite{ogilvie}
Suppose that we have a general {\em analytic} operator (including any inverse propagator) $G^{-1}(\lambda)$ that depends
on a parameter $\lambda$. Let $a_\alpha$ be a particular eigenvalue,
\begin{eqnarray}
\label{ggen}
G^{-1}(\lambda) \left|a_{\alpha}(\lambda) \right.\rangle=a_{\alpha}(\lambda) \left| a_{\alpha}(\lambda) \right.\rangle.
\end{eqnarray}
The secular equation for the eigenvalues of $G^{-1}$ is an analytic function in $\lambda$. We may thus replicate our earlier considerations to obtain similar results.
In doing so, we see that if $a_{\alpha}(\lambda)$ changes from being purely real to becoming complex as we vary the
parameter $\lambda$ beyond a particular threshold value $\lambda_*$
(i.e., if $a_{\alpha}(\lambda>\lambda_*)$ is real and $a_{\alpha}(\lambda<\lambda_*)$ is complex,
or the vice versa), then the imaginary
part of $a_{\alpha}(\lambda)$ will scale (for $\lambda<\lambda_*$ in the first
case noted above and for $\lambda>\lambda_*$ in the second one) as
$ %\begin{eqnarray}
\mbox{Im } \{a_{\alpha}(\lambda)\} \propto|\lambda-\lambda_*|^{1/2}.
$ %\end{eqnarray}
A particular such realization is associated with the spectrum of
a non-Hermitian Hamiltonian [playing the role of $G^{-1}$ in Eq. (\ref{ggen})]
which, albeit being non-Hermitian, may have real eigenvalues
(as in ${\cal PT}$ symmetric Hamiltonians).\cite{PT}
In this case, the crossover occurs when a system becomes ${\cal PT}$
symmetric as a parameter $\lambda$ crosses a threshold $\lambda_*$.

 Similarly, if $a_{\alpha}(\lambda)$ changes from being pure imaginary to complex at $\lambda=\lambda_*$,
then the real part of the eigenvalue will
scale in the same way. That is, in the latter instance,
$ %\begin{eqnarray}
\mbox{Re } \{a_{\alpha}(\lambda)\}  \propto|\lambda-\lambda_*|^{1/2}.
$ %\end{eqnarray}

\begin{figure}
  \begin{center}
    \vspace{.2in}
%    \hspace{.1in}
\includegraphics[width=.8\columnwidth]{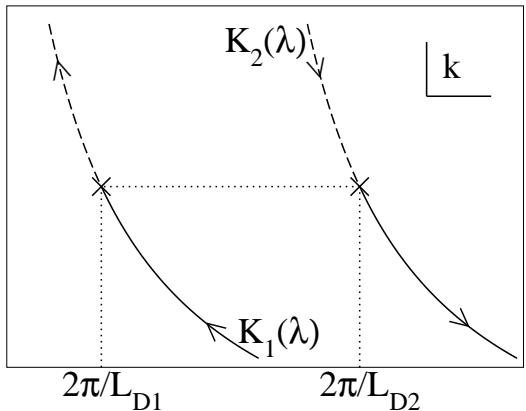}
    \caption{
``Jumps'' in the modulation length: The figure shows the evolution of
the poles associated with two different eigenvectors %[in
                                %Eq. (\ref{ggen})]
 with the
parameter $\lambda$ in the complex $k$-plane. The solid portion of the trajectories show which
pole corresponds to the dominant term (larger correlation length) in the correlation function. The $\times$-s denote the
poles at $\lambda=\lambda_*$ and the arrows denote the direction of
increasing $\lambda$. It is evident, therefore, that the modulation length corresponding to the
dominant term jumps from $L_{D1}$ to $L_{D2}$ as $\lambda$ crosses the
threshold value $\lambda_*$.
}\label{modlengthjumps}
  \end{center}
\end{figure}

Our next brief remark pertains to some theories with multi-component
fields, e.g. $n$ component theories with Hamiltonians of the form,\cite{ref:nussinovAPT}
\begin{eqnarray}
H=\frac{1}{2N}\sum_{\vec{k},i,j}v_{ij}(k)s_i(\vec{k})s_j(\vec{k}),
\label{Hamkspaceirrot}
\end{eqnarray}
in which, unlike Eq. (\ref{Hamkspace}) (as well as standard $O(n)$ theories), the interaction kernel $v_{ij}$ might not be 
diagonal in the internal field indices $i,j=1,2, \ldots, n$.
An example is afforded by a field theory in which $n$ component fields
are coupled minimally to a spatially uniform (and thus translationally
invariant) non-Abelian gauge background which emulates a curved space
metric.\cite{ref:nussinovAPT} In this case, the index $\alpha$ in
Eq. (\ref{ggen}) is a composite of an internal field component
coordinate $w=1,2, \ldots, n$ and $\vec{k}$-space coordinates.
For each of the $n$ branches $w$, we may determine the associated
$\vec{k}$-space zero eigenvalue of Eq. (\ref{ggen}) which we label by
$K_{w}$ (i.e., $a_{w,k=K_{w}}(\lambda) =0$).
The largest correlation is length is associated
with the eigenvector which exhibits the smallest value of $|\mbox{Im } K_{w}|$.
As usual, as $\lambda$ is varied, we may track for each of the $n$ branches, the trajectories
poles of $G$ in the complex $k$-plane.
Although the location of the multiple poles may vary continuously with the parameter $\lambda$, the dominant poles 
(those associated with the largest correlation length) might discontinuously change
from one particular subset of eigenvectors to another (see
Fig. \ref{modlengthjumps}). As such, the correlation function of the system may show jumps in its
dominant modulation length at large distances as $\lambda$ crosses a threshold value $\lambda_*$ even though no
transitions (nor cross-overs similar to that of Fig. (\ref{cixpoles})
which form the focus of this work) are occurring.
Such jumps in the large distance modulation lengths appear in
$O(n)$ systems defined on a fixed, translationally invariant,
non-Abelian background or metric as in Ref. \cite{ref:nussinovAPT}.

In Appendix \ref{secfermi}, we discuss exponents associated with
lock-ins of correlation and modulation lengths in Fermi systems. When
dealing with zero temperature behavior, we use the chemical potential
$\mu$ as the control parameter $\lambda$. We discuss metal-insulator
transition, exponents in Dirac systems and topological
insulators. Additionally, we comment on crossovers related to changes in the Fermi surface
topology as well as those related to situations with divergent effective mass.

%%%%%%%%%%%%%%%%%%%%%%%%%%%%%%%%%%%%%%%%%%%%%%%%%%%%%%%%%%%%
%\secn{Implications for the time domain: Josephson time scales {\textcolor{red}{and resonance lifetimes}}}
\secn{Implications for the time domain: Josephson time scales and resonance lifetimes}
%%%%%%%%%%%%%%%%%%%%%%%%%%%%%%%%%%%%%%%%%%%%%%%%%%%%%%%%%%%%
As we alluded to above, the results that we derived earlier that pertained to length scales can also be
applied to time scales in which case we look at a temporal correlation
function characterized by decay times (corresponding to correlation
lengths) and  oscillation periods (corresponding to modulation lengths).
We may obtain decay time and oscillation period exponents
whenever one of these time scales freezes to a constant value as some
parameter $\lambda$ crosses a threshold value $\lambda_*$.

Many other aspects associated with length scales
have analogs in the temporal regime. Towards this end, in what
follows, we advance the notion of a ``Josephson time  scale''. We first very briefly review below the concept of a Josephson length scale.
In many systems [with correlation functions similar to
Eq. (\ref{anomr})], just below the critical temperature, 
the correlation function as a function of wave-vector, $k$ 
behaves as %switches from from canonical ($\eta=0$) to anomalous ($\eta\neq0$) behavior as $\{q:0\to\infty\}$, 
\begin{eqnarray}
G(k)\propto\left\{\begin{array}{ll}
k^{-2+\eta}&\mbox{for}~k\gg1/\xi_J,
\\
k^{-2}&\mbox{for}~k\ll1/\xi_J,
\end{array}
\right.
\end{eqnarray}
thus defining the Josephson length scale, 
$\xi_J$.\cite{josephson}
Such an argument may be extended to a time scale, $\tau_J$ (real or imaginary) 
in systems with Lorentz invariant propagators. For a given wave-vector
$k$, $\tau_J$ may be  defined as,
\begin{eqnarray}
G(k,\omega)\propto\left\{\begin{array}{ll}
\omega^{-2+\eta_t}&\mbox{for}~\omega\gg1/\tau_J,
\\
\omega^{-2}&\mbox{for}~\omega\ll1/\tau_J,
\end{array}
\right.
\end{eqnarray}
where $\omega$ is the frequency conjugate to time while performing the Fourier transform
and $\eta_t(\neq0)$ is an anomalous exponent for the time variable.

We next briefly allude to another possible 
simple application of our result. As is well known in high energy
(see, e.g., Ref. \cite{peskin} for a standard textbook treatment) and many
body theories, the Fourier transform of the two two point correlation function 
$G(\vec{k}, \omega)$ generally exhibits isolated poles corresponding to the one particle states as well as bound
states and a branch cut that reflects a continuum of multi-particle states (i.e., two particles or more).
Such a continuum of states arises when the squared four-momentum
$p^2\equiv E^2/c^2-\vec{p}^2$ exceeds the threshold necessary
for the production of two particles, i.e., $p^{2} \ge (2m)^{2}c^2$ with
$m$ the particle rest mass and $c$ the speed of light. Single particle (and bound) states
and continuous multi-particle 
states lead to the aforementioned respective single poles and branch
cuts along the real $p^{2}$ axis. We may consider an application of
our ideas in the vicinity of zero energy bound states  
(as in, e.g., the Feshbach resonance of the BCS to BEC
crossover,\cite{regalgreinerjin, zwierlein, hodiener, nussinovs}
% e.g., CITE C. A. Regal, M. Greiner and D. S. Jin, Phys. Rev. Lett. {\bf 92}, 040403 (2004); 
% M. Zwierlein, et.al., Phys. Rev. Lett. {\bf 92}, 120403 (2004);
% R. B. Diener and T-L. Ho, Phys. Rev. Lett. {\bf 94}, 090402 (2005)
% Z. Nussinov and S. Nussinov, Physical Review A {\bf 74}, 053622 (2006) 
in dilute gases where the crossover is driven by varying an attractive contact interaction of strength 
$g$) when 
poles on the real axis are just about to splinter into poles with a infinitesimal imaginary part. Generally, when, by virtue of self-energy corrections, the poles attain a finite imaginary part in the $p^{2}$ plane,
the corresponding states attain a finite lifetime (with the lifetime being the analog of the correlation length/time in the two-point correlation
functions that we discussed hitherto). The relations (and exponents)
that we derived thus far may be applied, {\it mutatis mutandis}, for
the description of processes associated with the depinning of the
poles off the real axis, due to the imaginary part of the self energy $\Sigma$,
leading to resonances with a finite life-time. This relates to the scaling of the lifetime
$\tau$ 
of resonances in cold atomic gases as a function of $(g_{0}-g)$ where
$g_{0}$ is the strength of the contact interaction at the BCS to BEC
crossover point.

%%%%%%%%%%%%%%%%%%%%%%%%%%%%%%%%%%%%%%%%%%%%%%%%%%%%%%%%%%%%%%%%%%%%%%%%%%%%%%%%
\secn{Chaos and glassiness}\label{secchaos}
%%%%%%%%%%%%%%%%%%%%%%%%%%%%%%%%%%%%%%%%%%%%%%%%%%%%%%%%%%%%%%%%%%%%%%%%%%%%%%%%
%\textcolor{rred}{
Thus far, we have considered phases in which the
modulation length is well defined. For completeness, in this section,
we mention situations in which aperiodic phases may be observed. The general possibility of such
phenomena in diverse arenas is well known.\cite{bakreview,glasschaos} We focus here on translationally invariant systems
of the form of Eqs. (\ref{Ham},\ref{Ham_soft}) with competing interactions on different scales that lead to kernels such as 
\begin{eqnarray}
v(k)=k^4-c_1k^2+c_2,
\label{vkchaos2}
\end{eqnarray}
where $c_1$ and $c_2$ are positive constants may give rise to glassy structures for non
zero $u$. 
Such a dispersion may arise in the continuum (or small $k$) limit of hyper-cubic
lattice systems with next nearest neighbor interactions (giving rise to the $k^4$
term) and nearest neighbor interactions (giving rise to the $k^2$ term).
Within replica type approximations, such kernels that have a finite 
$k$ minimum  (i.e., ones with $c_{1}>0$) may lead to extensive configurational entropy that might enable extremely slow
dynamics.\cite{ref:nussinovAPT, schmalianwolyneschaos}

The simple key idea regarding ``spatial chaos'' is as follows. It is well known that nonlinear dynamical
systems may have solutions that exhibit chaos. This has been extensively applied in the time
domain yet, formally, the differential equations governing the system may determine
not how the system evolves as a function of the time $t$ 
but rather how fields change as a spatial coordinate ($x$) [replacing the time ($t$)].
Under such a simple swap of $t \leftrightarrow x$, we may observe
spatial chaos as a function of the spatial coordinate $x$. 
In general, of course, more than one coordinate may be involved.
The resultant spatial configurations may naturally correspond to amorphous systems and
realize models of structural glasses. 
% Such a correspondence has been discussed in the realm of spin
% glasses, wherein spin glass transitions in random Potts systems and
% hard computational problems coincide with
% transitions from ordered to chaotic phases in derived mechanical
% systems.
A related correspondence in disordered systems has been found in random Potts systems wherein
spin glass transitions coincide with transitions from regular to
chaotic phases in derived dynamical analogs.\cite{dandan}

In the translationally invariant systems that form the focus of our study, an effective free energy of the form
\begin{eqnarray}
{\cal
  F}[s]&=&\frac{1}{2}\int\frac{d^dk}{(2\pi)^d}(v(\vec{k})+\mu)|s(\vec{k})|^2+\nonumber\\
&&~~~~~~~~~\frac{u}{4}\int
d^dx (S^2(\vec{x})-1)^2 \label{freenrg}
\end{eqnarray}
is generally associated with single component ($n=1$) systems of the form of Eqs. ( \ref{Ham_soft}).
In Eq. (\ref{freenrg}), $\mu$ represents the deviation from the transition temperature in Ginzburg-Landau theories
(or equivalently related to Eq. (\ref{gmt})).

Euler-Lagrange equations for the spins $S(\vec{x})$ are obtained by
extremizing the free energy of Eq. (\ref{freenrg}). These equations are, generally,
 nonlinear differential equations (as
discussed in Appendix \ref{appeullag}). As is well appreciated, however,
nonlinear dynamical systems may exhibit chaotic behavior. In
general, a dynamical system may, in the long time limit, either veer towards a
fixed point, a limit cycle, or exhibit chaotic behavior.
We should therefore expect to see such behavior in the spatial
variables in systems which are governed by Euler-Lagrange equations
with forms similar to nonlinear dynamical systems.
Upon formally replacing the temporal coordinate by a spatial coordinate,
chaotic dynamics in the temporal
regime map onto to a spatial amorphous (glassy) structure.

\begin{figure}
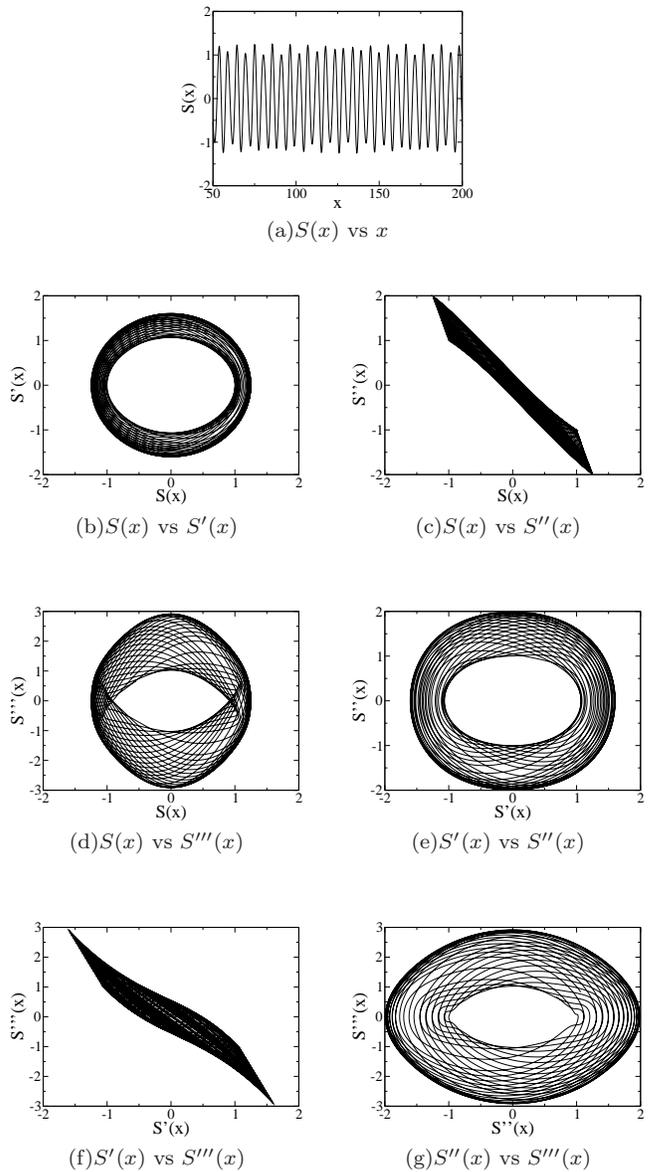

  \begin{center}
%    \vspace{.2in}
%    \hspace{-1in}
\vspace{.1in}
\subfigure[$S(x)$ vs
  $x$]{\includegraphics[width=.45\columnwidth]{chaos2st.eps}\label{chaos2st}}
% \qquad
% \vspace{.1in}
% \subfigure[$|s(k)|$ vs
%   $k$]{\includegraphics[width=.45\columnwidth]{chaos2sk.eps}\label{chaos2sk}}
\\
\vspace{.1in}
\subfigure[$S(x)$ vs $S'(x)$]{\includegraphics[width=.45\columnwidth]{chaos2sds.eps}\label{chaos2sds}}\qquad
\vspace{.1in}
\subfigure[$S(x)$ vs $S''(x)$]{\includegraphics[width=.45\columnwidth]{chaos2sd2s.eps}\label{chaos2sd2s}}\\
\vspace{.1in}
\subfigure[$S(x)$ vs $S'''(x)$]{\includegraphics[width=.45\columnwidth]{chaos2sd3s.eps}\label{chaos2sd3s}}\qquad
\vspace{.1in}
\subfigure[$S'(x)$ vs $S''(x)$]{\includegraphics[width=.45\columnwidth]{chaos2dsd2s.eps}\label{chaos2dsd2s}}\\
\vspace{.1in}
\subfigure[$S'(x)$ vs $S'''(x)$]{\includegraphics[width=.45\columnwidth]{chaos2dsd3s.eps}\label{chaos2dsd3s}}\qquad
\vspace{.1in}
\subfigure[$S''(x)$ vs $S'''(x)$]{\includegraphics[width=.45\columnwidth]{chaos2d2sd3s.eps}\label{chaos2d2sd3s}}\\
    \caption{
Glassiness in system with $v(k)$ as in Eq. (\ref{vkchaos2}) with $c_1=5$, 
$c_2=4$ and $u=1$ and $\mu=1$ in Eq. (\ref{freenrg}). 
% \subref{svx}:   
% \subref{dsvs}: 
% \subref{d2svds}: 
% \subref{svd2s}: 
}
    \label{chaos2}
  \end{center}
\end{figure}

% The glassiness is shown in Fig. \ref{chaos2} where the chaotic
% attractor for this system has been plotted.
In Fig. \ref{chaos2st}, we illustrate the spatial amorphous glass-like chaotic behavior that a one-dimensional rendition of the system of Eq. (\ref{vkchaos2}) exhibits.
In Figs. \ref{chaos2sds}--\ref{chaos2d2sd3s}, we provide plots of the spatial derivatives of different order vs each other (and $S(x)$ itself).

\begin{figure}
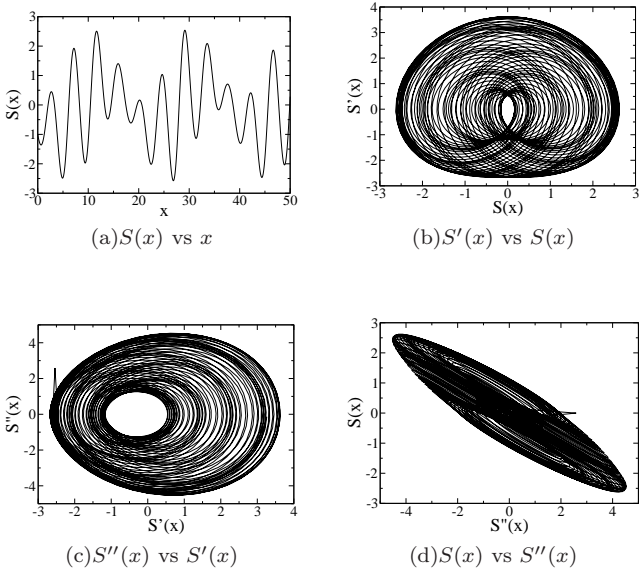

  \begin{center}
%    \vspace{.2in}
%    \hspace{-1in}
\vspace{.1in}
\subfigure[$S(x)$ vs $x$]{\includegraphics[width=.45\columnwidth]{jerksxvsx.eps}\label{svx}}\qquad
\vspace{.1in}
\subfigure[$S'(x)$ vs $S(x)$]{\includegraphics[width=.45\columnwidth]{jerkdsvssx.eps}\label{dsvs}}\\
\vspace{.1in}
\subfigure[$S''(x)$ vs $S'(x)$]{\includegraphics[width=.45\columnwidth]{jerkd2svsdsx.eps}\label{d2svds}}\qquad
\vspace{.1in}
\subfigure[$S(x)$ vs $S''(x)$]{\includegraphics[width=.45\columnwidth]{jerksvsd2sx.eps}\label{svd2s}}
    \caption{
Example of aperiodic structure inspired by system with nonlinear
jerks. Here $J(S(x),S'(x),S''(x))=-2S'(x)+(|S(x)|-1)$ and initial
conditions are $S(0)=-1$, $S'(0)=-1$, $S''(0)=1$ (chosen from
Ref. \cite{sprottsimplechaosajp}).
% \subref{svx}:   
% \subref{dsvs}: 
% \subref{d2svds}: 
% \subref{svd2s}: 
}
    \label{jerkex}
  \end{center}
\end{figure}

Another example comes from the spatial analog of dynamical
systems with nonlinear ``jerks''. It is well known that systems with nonlinear ``jerks'' often give rise
to chaos\cite{sprottsimplechaosajp} ``Jerk'' here refers to the time
derivative of a force, or, something which results in a change in the
acceleration of a body.
Translating this idea from the temporal regime to the spatial regime,
one can expect to obtain a aperiodic/glassy structure in a system for
which the Euler Lagrange equation, Eq. (\ref{eullag}) may seem simple.
For example, if we have the following, Euler Lagrange equation for a
particular one-dimensional system,
\begin{eqnarray}
S'''(x)=J(S(x),S'(x),S''(x)),
\end{eqnarray}
with a non-linear function $J(S(x),S'(x),S''(x))$ then the system may have
aperiodic structure. An example is depicted in Fig. (\ref{jerkex}).

% %%%%%%%%%%%%%%%%%%%%%%%%%%%%%%%%%%%%%%%%%%%%%%%%%%
% \subsecnnn{Poly-spiral solutions of the Euler-Lagrange equations in $O(n\ge2)$ systems}
% %%%%%%%%%%%%%%%%%%%%%%%%%%%%%%%%%%%%%%%%%%%%%%%%%%
We now discuss $O(n)$ systems and illustrate the existence of
periodic solutions (and absence of chaos) in a broad class of systems.

The Euler-Lagrange equations for the system in Eq. (\ref{freenrg}) 
[written longhand in Eqs. (\ref{eullag}, \ref{eullaglat})]
become linear in case of ``hard'' spins, i.e., when the $O(n)$ condition is
strictly enforced, i.e., $u\to\infty$.
In this limit, all configurations in the system can be described by a
finite set of modulation wave-vectors (as was the case for the ground
states in Sec. \ref{ltc}).

There are several ways to discern this result. First, it may be simply argued
that since the Euler-Lagrange equations represent a {\em finite} set of
coupled {\em linear} ordinary differential equations, chaotic
solutions are not present. The configurations, therefore must be
characterized by a finite number of modulation wave-vectors.

A second approach is more quantitative. 
The idea used here is the same as the one used in
Ref. \cite{zohar_com}. An identical construct can be applied to illustrate
that spiral/poly-spiral states are the only possible states that
satisfy the Euler-Lagrange equation if $n>1$.
With $v$ being a functional of the lattice Laplacian of Eq. (\ref{LLaplace}), the lattice rendition of the Euler-Lagrange equations in Fourier space reads
\begin{eqnarray}
D(\Delta_{\vec{k}})s(\vec{k})=0.\label{eullaglatkuinf}
\end{eqnarray}
In what follows we consider what transpires when the Euler-Lagrange equations
have real wave-vectors ${\cal K}=\{\vec{q}_m\}$vas solutions. 
\begin{eqnarray}
\left.D(\Delta_{\vec{k}})s(\vec{k})\right|_{\vec{k}=\vec{q}_m}=0.\label{eullaglatkuinfqm}
\end{eqnarray}
To obtain a bound on the number of wave-vectors that can be
used to describe a general configuration satisfying the Euler-Lagrange
equations, we consider general situations wherein 
{\bf (i)} $2(\vec{q}_m \pm \vec{q}_{m'}) \neq \vec{k}_{rec}$ for any $\vec{q}_m,\vec{q}_{m'}\in{\cal K}$, where $\vec{k}_{rec}$ represents a reciprocal lattice
vector; and, 
{\bf (ii)} $\vec{q}_m \pm \vec{q}_{m'} \neq \vec{q}_p \pm \vec{q}_{p'}$ for any $\vec{q}_m,\vec{q}_{m'},\vec{q}_p,\vec{q}_{p'}\in{\cal K}$.
Let a particular state be described as
\begin{eqnarray}
\vec{S}_0(\vec{x})=\sum_m\vec{a}_m e^{-i\vec{q}_m\cdot\vec{x}},
\end{eqnarray}
where the vectors $\vec{a}_m$ have $n$ components for $O(n)$ systems.
As the states must have real components, the above equation must
take the form,
\begin{eqnarray}
\vec{S}_0(\vec{x})=\sum_{m=1}^{N_q} \left(\vec{a}_m e^{-i\vec{q}_m\cdot\vec{x}} + \vec{a}_m^* e^{i\vec{q}_m\cdot\vec{x}}\right).\label{s0real}
\end{eqnarray}
In the above, $\vec{a}_m^*$ denotes the vector whose components are complex
conjugate those of the vector $\vec{a}_m$. In Eq. (\ref{s0real}), we do not
count terms involving the wave-vectors $\vec{q}_m$ and $-\vec{q}_m$ separately as such terms
has been explicitly written in the sum.

We next define the complex vectors $\{\vec{U}_m\}$ and $\{\vec{V}_m\}$ as
%$\{\ket{U_m}\}$ and  $\{\ket{V_m}\}$ as,
\begin{eqnarray}
\vec{U}_m=\vec{a}_me^{-i\vec{q}_m\cdot\vec{x}},\nonumber\\
\vec{V}_m=\vec{a}_me^{i\vec{q}_m\cdot\vec{x}}.
\end{eqnarray}
The $O(n)$ normalization condition can then be expressed as,
\begin{eqnarray}
\sum_{m} |\vec{U}_m|^{2}&=&n,\nonumber\\
\sum_{m}|\vec{V}_m|^{2}&=&n,\nonumber\\
\sum_{\vec{q}_m-\vec{q}_{m'}=\vec{A}} \left(\vec{U}^{*}_m \cdot \vec{U}_{m'}\right.&+&\left.\vec{V}^{*}_{m'} \cdot\vec{V}_{m}\right)+\nonumber\\
\sum_{\vec{q}_m+\vec{q}_{m'}=\vec{A}} \left(\vec{U}^{*}_m  \cdot \vec{V}_{m'}\right.&+&\left.\vec{U}^{*}_{m'} \cdot \vec{V}_{m}\right)=0.\label{uvon}
\end{eqnarray}
Solutions to Eq. (\ref{uvon}) are spanned by the set of mutually
orthonormal basis vectors $\{\vec{U}_{m}\}\cup\{\vec{V}_{m}\}$. As
these $2N_q$ basis vectors are described by $n$-components each,
it follows that
\begin{eqnarray}
N_q\le n/2.
\end{eqnarray}
Therefore, such states satisfying the Euler-Lagrange equations for an
$O(n\ge2)$ system can at most be characterized by $n/2$ pairs of wave-vectors.
These states can be described by $N_q$ spirals (or ``poly-spirals'') each of which is described in a
different orthogonal plane.

A few remarks are in order.
\begin{itemize}
% \subsecnnn{Soft-spins on the lattice with Euler-Lagrange equation
%   having real wave-vector solutions}%Lattice, finite $u$ (real wave-vector zeros of $D(\Delta_{\vec{q}})=0)$}
%-------------------------------------------------------------------------------
\item
When $u$ in Eq. (\ref{freenrg}) is finite, i.e., in the soft spin regime, poly-spiral solutions could be present even though
aperiodic solutions are also allowed.
\item
{\em Continuum limit:}
%($D(\Delta_{\vec{q}})=0$):
%------------------------------------------------
In the hard-spin limit, i.e., $u\to\infty$ in Eq. (\ref{freenrg}), if the Fourier space
Euler-Lagrange equation is satisfied by non-zero real wave-vectors, we
have poly-spiral solutions as in the lattice case.
When $u$ is finite, aperiodic solutions may also be present.
\item
% \subsecnnn{Fourier space Euler-Lagrange equation without real
%   wave-vector solutions}
% Zeros of $D$ that do not correspond to a value attainable for $\Delta_{\vec{q}}$
% with a real-wave vector $\vec{q}$
% ----------------------------------------------
If the Fourier space Euler-Lagrange equation does not have any real
wave-vector solution, poly-spiral states are not observed.
\end{itemize}

%\secn{Exponent observed in chaos via intermittency}
In nonlinear dynamical systems, chaos is often observed via
{\em intermittent phases}. As a tuning parameter $\lambda$ is varied, the
system enters a phase in which it jumps between periodic and
aperiodic phases until the length of the aperiodic phase diverges. This
divergence is characterized by an exponent $\nu=1/2$ similar to ours.\cite{pomeauintermitentexponent}

%%%%%%%%%%%%%%%%%%%%%
\secn{Conclusions}
%%%%%%%%%%%%%%%%%%%%%
\label{conc}
Most of the work concerning properties of the correlation functions
in diverse arenas, has to date focused on the correlation lengths
and their behavior. In this work, we examined the oscillatory character
of the correlation functions when they appear.

We furthermore discussed when viable non-oscillatory 
spatially chaotic patterns may (or may not appear); in these,
neither uniform nor oscillatory behavior is found. Our results are 
universal and may have many realizations. Below, we provide a brief 
synopsis of our central results.

\begin{enumerate}
\item
We have shown the existence of a universal modulation length exponent
$\nu_L=1/2$ [Eq. (\ref{dle})]. Here the scaling could be as a function of some general
parameter $\lambda$ such as temperature.
This is observed in systems with analytic crossovers
including the commensurate-incommensurate crossover in the ANNNI
model.
\item
In certain situations the above exponent could take other rational
values [Eq. \ref{traj}].
\item
This result also applies to situations where a correlation length may
lock in to a constant value as the parameter $\lambda$ is varied
across a threshold value [as in Eq. (\ref{xitst})].
\item
We extended our result to include situations in which the crossover
might take place at a branch point. In this case irrational exponents
could also be present.
In Eqs. (\ref{nulanom}, \ref{nucanom}), we provide universal scaling
relations for correlation and modulation lengths.
\item
We illustrate that discontinuous jumps in the modulation/correlation
lengths mandate a thermodynamic phase transition.
\item
We showed that in translationally invariant systems (with rotational and/or reflection
symmetry), the total number of correlation and modulation lengths 
is generally conserved as the general parameter $\lambda$ is varied.
\item
Our results apply to both length scales as well as time scales. We further introduce the notion of a
Josephson time scale.
\item
We comment on the presence of aperiodic modulations/amorphous states
in systems governed by nonlinear Euler-Lagrange equations.
We illustrate that in a broad class of multi-component systems chaotic
phases do not arise. Spiral/poly-spiral solutions appear instead.
\item
Our results have numerous applications. We discussed several non-trivial consequences for classical
system in the text. For completeness, in Appendix \ref{secfermi}, we discuss, rather simple applications of our results
to non-interacting Fermi systems. We mention situations in
which the Fermi surface changes topology, situations
with divergent effective masses and the metal-insulator
transition.
We further discuss applications to many other systems including Dirac systems and
topological insulators.
Aside from uniform
and regular modulated periodic states of various
strongly correlated electronic systems,\cite{salamonmanganites, pnic1, pnic2, tran, cuprate1, cuprate2,
  cuprate3, cuprate4, steve} there are numerous
suggestions and indications of glassy (and spatially
non-uniform or chaotic) behavior that naturally
lead to high entropy in these systems, e.g., see, e.g., Refs. \cite{schmalianwolyneschaos,
   dielectric, vladglasschaos, zoharbalatsky, clarkentropyparadox}. 
When spatial modulations are present in the ground states of
rotationally invariant (and other) systems, they may lead to
``holographic''-like entropy (as in large $n$ renditions), \cite{ref:nussinovAPT}.
In future work,
 we will elaborate on non-trivial consequences of our results for
 interacting Fermi systems. 
\end{enumerate}

Our general analysis regarding the expansion of the inverse correlator
$G^{-1}$ as a function of $k$  about points $k_*$ and the myriad
conclusions that we draw from it (including exponents) may, in some
cases, be viewed as a formal analog of Ginzburg-Landau method of
expanding an effective free energy ${\cal{F}}$ in an order parameter
field $\phi$ (i.e., $\delta k \leftrightarrow \phi$ and $G^{-1}
\leftrightarrow {\cal{F}}$).

Finally, we make a brief parenthetic  
remark concerning the ``fractal dimension'' in glasses and other
systems. The notion of fractal
dimensionality was recently applied in Ref. \cite{ma_fractal} 
based on a comparison between the atomic volume and the reciprocal of the dominant 
peak $K_{R}$ in the structure factor in metallic
glasses. Specifically, the volume $V \sim K_{R}^{-D_{f}}$ with $D_{f}$
being the fractal dimension. This definition is very intuitive and
such a relation between volume and structure factor peaks is to be
expected for a system of dimension $D_{f}$ if all natural scales in
the parameter expand or contract with temperature (or other
parameters) in unison. However, as we elaborated on at length, aside
from global changes in the lattice constant, $K_{R}$ can change
non-trivially with temperature and other paramters in some regular
lattice and other systems. Formally, this may give rise to an
effective non-trivial fractal dimension in various systems.

{\bf Acknowledgments.}
The work at Washington University in St Louis has been supported by the National Science Foundation
under NSF Grant numbers DMR-1106293 (Z.N.) and DMR-0907793 (A.S.) and
by the Center for Materials Innovation.
Z.N.'s research at the KITP was supported, in part, by the NSF under Grant No. NSF PHY11-25915. Z.N. is grateful to
the inspiring KITP workshops on ``Electron glass'' and ``Emerging concepts in glass physics'' in the summer of 2010. 
V.D. was supported by the NSF through Grant DMR-1005751.

\begin{appendix}

%%%%%%%%%%%%%%%%%%%%%%%%%%%%%%%%%%%%%%%%%%%%%%%%%%%%%%%%%%%%
\secn{Fermi systems}\label{secfermi}
%%%%%%%%%%%%%%%%%%%%%%%%%%%%%%%%%%%%%%%%%%%%%%%%%%%%%%%%%%%%
In this appendix, we discuss several examples of non-interacting fermionic systems
where we observe a correlation or modulation length exponent.
We will, in what follows, ignore spin degrees of freedom which lead to
simple degeneracy factors for the systems that we analyze.
In non-interacting Fermi systems, the mode occupancies are given by the
Fermi function. That is,
\begin{eqnarray}
\langle n(\vec{k}) \rangle=\langle c^{\dagger}(\vec{k}) c(\vec{k}) \rangle=\frac{1}{e^{\beta(\epsilon(\vec{k})-\mu)}+1},
\end{eqnarray}
where $c(\vec{k})$ and $c^{\dagger}(\vec{k})$ are the annihilation and
creation operators at momentum $\vec{k}$ and $\beta=1/(k_BT)$ with $T$
the temperature. % and $k_B$ the Boltzmann's constant.
The correlation function associated with the amplitude for hopping
from the origin to lattice site $\vec{x}$ is given by
\begin{eqnarray}
G(\vec{x})=\langle C^\dagger(0)C(\vec{x})\rangle
=\sum_{\vec{k}} \langle n(\vec{k}) \rangle e^{-i\vec{k}\cdot\vec{x}}.
\label{gxfermi}
\end{eqnarray}

Thus far, in most explicit examples that we considered we discussed
scaling with respect to a crossover temperature. In what follows, we
will, on several occasions, further consider the scaling of
correlation and modulation lengths with the chemical potential
$\mu$. We will use the letter $\upsilon$ to represent exponents
corresponding to scaling with respect to $\mu$ and continue to use
$\nu$ to represent scaling with respect to the temperature $T$.

The existence of modulated electronic phases is well
known.\cite{salamonmanganites,pnic1, pnic2,tran, cuprate1,
  cuprate2,cuprate3, cuprate4, steve,qhstr1, qhstr2,
  qhstr3,jan,rmp_steve,gruner_rmp,gruner_book} In particular, the
Fermi wave-vector dominated response of diverse modulated systems as
evident in Lindhard functions, particular features of charge and spin
density waves dominated by Fermi surface considerations in quasi- one
dimensional and other systems have long been discussed and have
numerous experimental realizations in diverse
compounds.\cite{gruner_rmp,gruner_book} 
The exponents that we derived in this work appear for all electronic
and other systems in which a crossover occurs in the form of the
modulations seen in charge, spin, or other degrees of freedom. 
Our derived results concerning scaling apply to general
interacting systems. To highlight essential physics as it pertains
to the change of modulations in systems of practical importance, we
briefly review and further discuss free electron systems.

%%%%%%%%%%%%%%%%%%%%%%%%%%%%%%%%%%%%%%%%%%%%%%%%%%
\subsecn{Zero temperature length scales -- Scaling as a function of the
  chemical potential $\mu$}\label{ltl}
%%%%%%%%%%%%%%%%%%%%%%%%%%%%%%%%%%%%%%%%%%%%%%%%%%
We first consider a non-interacting fermionic system with a dispersion $\epsilon(\vec{k})$.
At zero temperature, the number of particles occupying the Fourier
mode $\vec{k}$ is given by
\begin{eqnarray}
\langle n(\vec{k}) \rangle
=
\left\{
\begin{array}{ll}
1&\mbox{for }\epsilon(\vec{k})<\mu
\\
0&\mbox{for }\epsilon(\vec{k})>\mu.
\end{array}
\right.
\end{eqnarray}
All correlation functions
as all other zero temperature thermodynamic properties, are determined by the Fermi
surface geometry. We now consider the correlation function of
Eq. (\ref{gxfermi}). This correlation function will generally exhibit
both correlation and modulation lengths. To obtain the modulation
lengths along a chosen direction
(the direction of the displacement $\vec{x}$), a ray along
that direction may be drawn. The intercept of this ray with the Fermi
surface provides the pertinent modulation wave-vectors.
As we vary $\mu$ we alter the density, $\rho$ via
\begin{eqnarray}
\rho=g_s\int_{\epsilon(\vec{k})<\mu}\frac{d^dk}{(2\pi)^d},
\end{eqnarray}
$g_s$ being the spin degeneracy ($g_s=2$ for non-interacting spin-half particles such as
electrons).
As the {\em Fermi surface topology} is varied, the following effects
may be observed.

\begin{figure*}
  \begin{center}
    \vspace{.2in}
%    \hspace{-1in}
\subfigure[]{\includegraphics[width=.65\columnwidth]{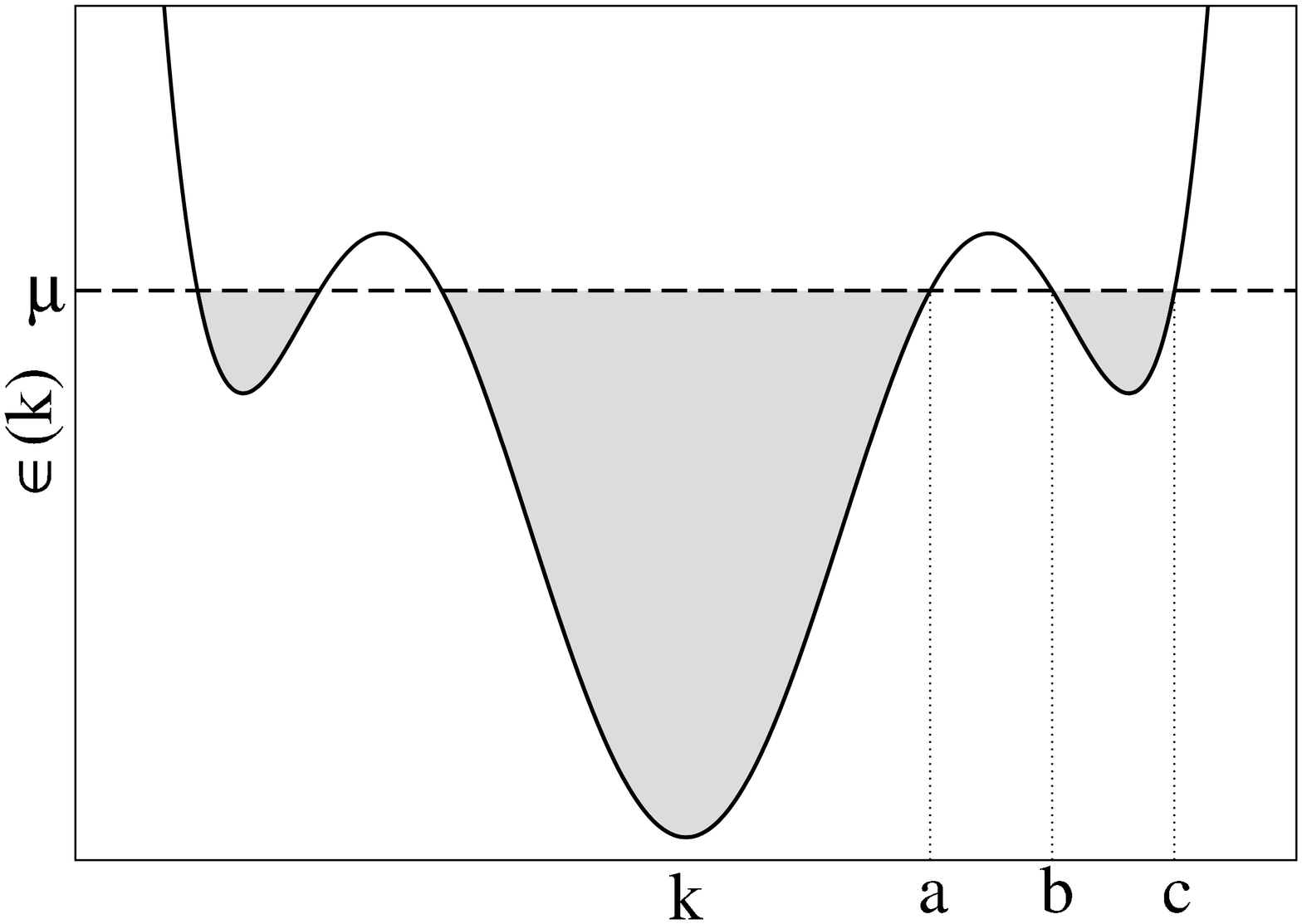}}\qquad
\hspace{.1in}
\subfigure[]{\includegraphics[width=.5\columnwidth]{circlefilled.eps}}\qquad
\hspace{.1in}
\subfigure[]{\includegraphics[width=.65\columnwidth]{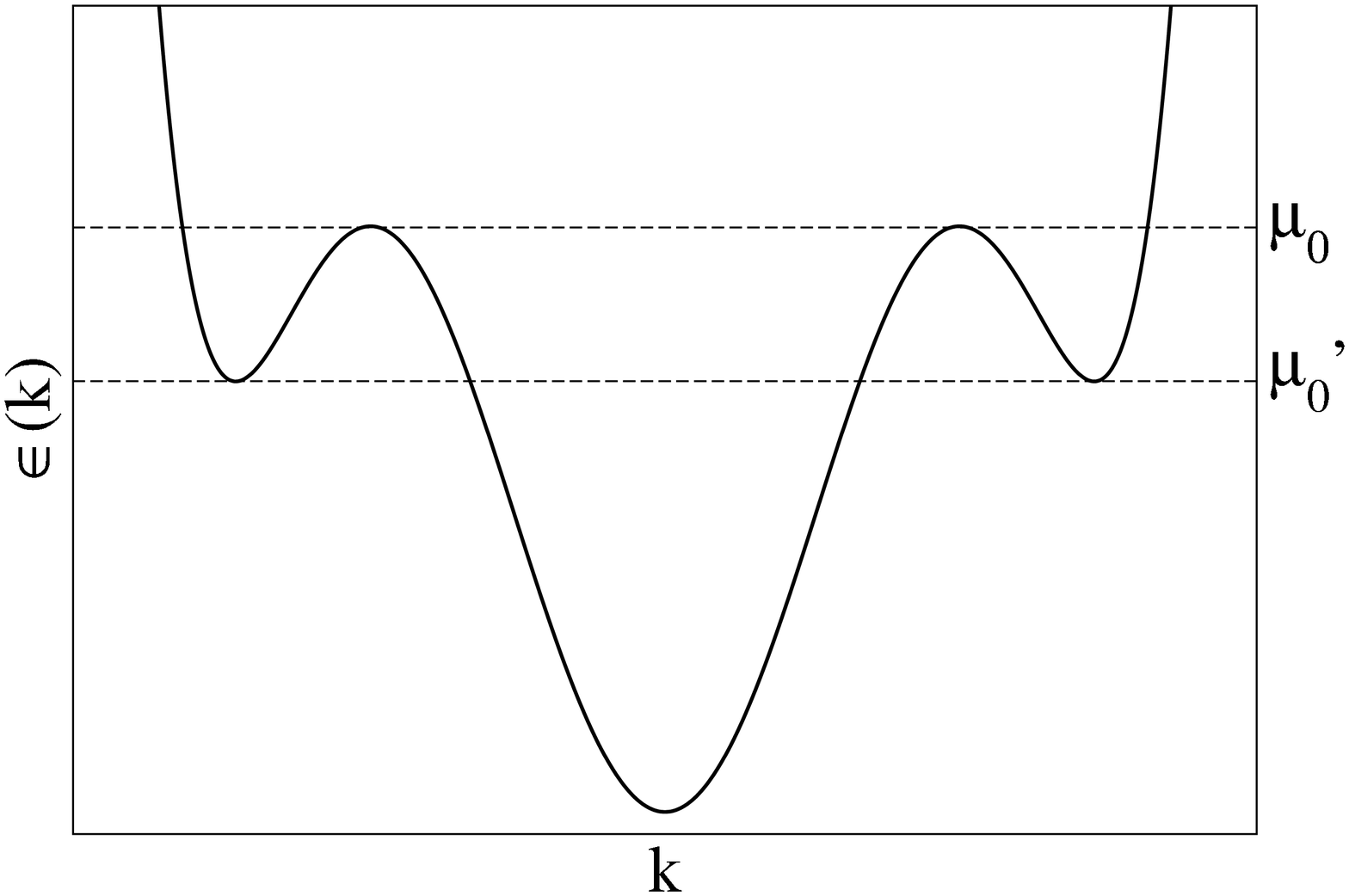}}
    \caption{Example of a Fermi system where the modulation length exponent is $1/2$. The gray region shows the filled states. When $\mu>\mu_0$, 
modulations corresponding to wave-vectors $k=a$ and $k=b$ cease to exist and we get an exponent of $1/2$ at this crossover. Similarly, when
$\mu<\mu_0'$, modulations corresponding to wave-vectors $k=b$ and $k=c$ die down.
}
    \label{ekvsmuexample}
  \end{center}
\end{figure*}

\begin{enumerate}
\item If two branches of the Fermi surface touch
  each other at $\mu=\mu_0$ and are disjoint for all other values of
  $\mu$, then a smooth crossover will appear from one set of modulation lengths to
 another with $|L_D-L_{D0}|\propto|\mu-\mu_0|$ on both sides of the
 crossover. This crossover will be associated with an exponent
 $\upsilon_L=1$ characterizing the scaling of the modulation lengths
 with deviations in the chemical potential.
An example where a crossover of this kind is realized is the $\epsilon_g=0$ case
of the schematic shown in Fig. \ref{metins} in which the crossover
occurs at $\mu=\mu_0$.
\begin{figure}[t]
  \begin{center}
%    \vspace{0.2in}
%    \hspace{-.1in}
    \includegraphics[width=\columnwidth]{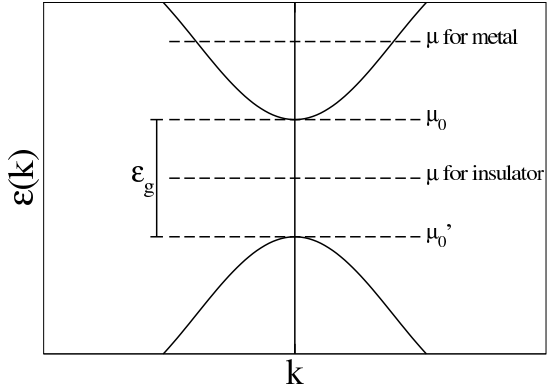}
    \caption{Transition from a metal to a band
      insulator. This figure is for illustration only.} 
    \label{metins}
  \end{center}
\end{figure}
Other examples of this occur at half filling of the square lattice tight
binding model and at three-quarters filling of the triangular lattice
tight binding model. These will be discussed later.
\item If on the other hand, one branch of the Fermi surface vanishes as we go past $\mu=\mu_0$, the crossover is not so smooth and we
get some rational fraction $\upsilon_L$ (usually $\upsilon_L=1/2$) as the crossover exponent: 
$|L_D-L_{D0}|\propto|\mu-\mu_0|^{\upsilon_L}$,  
on one side of the crossover.
An example of this is shown in Fig. \ref{ekvsmuexample}.
Here,
\begin{eqnarray}
|L_D-L_{D0}|=\frac{L_{D0}^2}{2\pi}\sqrt{\frac{2|\mu-\mu_0|}{|\epsilon''(2\pi/L_{D0})|}},
\end{eqnarray}
where $L_{D0}$ is the modulation length at the point where the $\mu=\mu_0$ line touches the $\epsilon(k)$ curve, such that $\epsilon'(2\pi/L_{D0})=0$
The hopping correlation function 
takes the form,
\begin{eqnarray}
G(x)&=&\frac{(ax)^{d/2}\mbox{J}_{d/2}(ax)}{(2\pi)^{d/2}x^d}-\frac{(bx)^{d/2}\mbox{J}_{d/2}(bx)}{(2\pi)^{d/2}x^d}\nonumber\\
&&+\frac{(cx)^{d/2}\mbox{J}_{d/2}(cx)}{(2\pi)^{d/2}x^d}\label{gxring},
\end{eqnarray}
where $\mu_0'<\mu<\mu_0$ and 
$a$, $b$ and $c$ in Eq. (\ref{gxring})
(corresponding to modulation lengths of $2\pi/a$,  $2\pi/b$ and $2\pi/c$)
are the values of $k$ for which $\epsilon(k)=\mu$ (as shown in Fig. \ref{ekvsmuexample}).
\end{enumerate}

\begin{figure}
  \begin{center}
%    \vspace{.2in}
%    \hspace{-1in}
\includegraphics[width=\columnwidth]{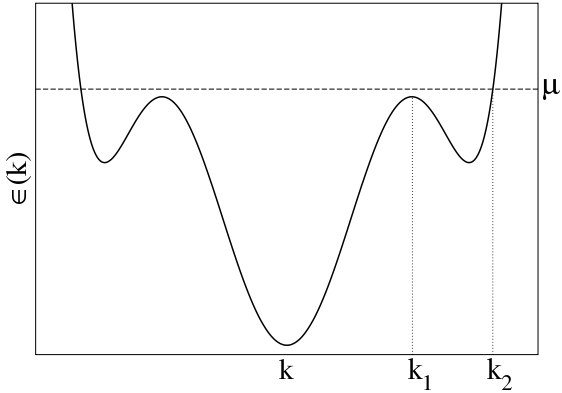}
    \caption{The same Fermi system as in Fig. \ref{ekvsmuexample}, but
      now with a chemical potential $\mu=\mu_0+\Delta$, slightly higher than $\mu_0$. The temperature is small but finite.
}
    \label{smalltemp}
  \end{center}
\end{figure}

At arbitrarily small but finite temperatures, the correlation function
exhibits modulations of all possible wavelengths.
The prefactor multiplying a term with spatial modulations at
wave-vector $\vec{k}$ is the exponential of  $(-|\epsilon(\vec{k})-\mu|)$.
An illustrative example is provided in Fig. \ref{smalltemp}. 
Apart from the dominant zero temperature modulations, associated with
the wave-vector $k_2$ in Fig. \ref{smalltemp}, 
at finite temperature, there are additional  contributions from
wave-vectors for which $|\epsilon(k)-\mu|$ is small relative to $k_BT$.
Near $k_2$, we can assume $\epsilon(k)$ is linear such that $\epsilon(k)\approx\mu+(k-k_2)\epsilon'(k_2)$.
Similarly, near $k_1$,  $\epsilon(k)-\mu\approx-\Delta-(k-k_1)^2\epsilon''(k_1)/2$, where $\Delta=\mu-\mu_0$ (see Fig. \ref{smalltemp}).
For large $\beta$, both these contributions are highly localized around $k_2$ and $k_1$ respectively making the above approximations very good
and the Fourier transforming integrals easy to evaluate ($\langle n(\vec{k})\rangle$ taking exponential and Gaussian forms).
We have,
\begin{eqnarray}
G(x)&=&\frac{(k_2x)^{d/2}\mbox{J}_{d/2}(k_2x)}{(2\pi)^{d/2}x^d}-\frac{2 (k_2x)^{d/2}\mbox{J}_{d/2-1}(k_2x)}{(2\pi)^{d/2}\beta\epsilon'(k_2)x^{d-1}}\nonumber\\
&&+\frac{e^{-\beta\Delta}(k_1x)^{d/2}\mbox{J}_{d/2-1}(k_1x)}{(2\pi)^{\frac{d-1}{2}}\sqrt{\beta\epsilon''(k_1)}x^{d-1}},
\end{eqnarray}
where $\beta\to\infty$ and $\Delta\to0$, such that $\beta\Delta\to\infty$.

% \begin{figure*}
%   \begin{center}
% %    \vspace{.2in}
% %    \hspace{-1in}
% \subfigure[ $~\mu=-0.01$]{\includegraphics[width=.6\columnwidth]{hexfermi-.01.eps}}\qquad
% \subfigure[ $~\mu=0.01$]{\includegraphics[width=.6\columnwidth]{hexfermi.01.eps}}\qquad
% \subfigure[ $~\mu=-0.01$(blue), $\mu=0$(green) and $\mu=0.01$(red)]{\includegraphics[width=.6\columnwidth]{hexfermiall}}
%     \caption{Fermi surface for a two-dimensional system with 
% $\epsilon(k)=\cos[k_x/\sqrt{2}]\cos[(k_x/2+k_y\sqrt{3}/2)/\sqrt{2}]\cos[(k_x/2-k_y\sqrt{3}/2)/\sqrt{2}]$. 
% This demonstrates a smooth crossover from one set of Fermi surface branches to another as $\mu$ is changed across $\mu=0$.
% The points where the crossovers take place are $(0,\pm\pi\sqrt{2/3})$, $(\pm\pi/\sqrt{2},\pm\pi/\sqrt{6})$.
% The modulation length exponent for this crossover is $1$.}
%     \label{hexfermi}
%   \end{center}
% \end{figure*}

\begin{figure}[b]
  \begin{center}
    \vspace{.5in}
%    \hspace{2in}
    \includegraphics[width=.8\columnwidth]{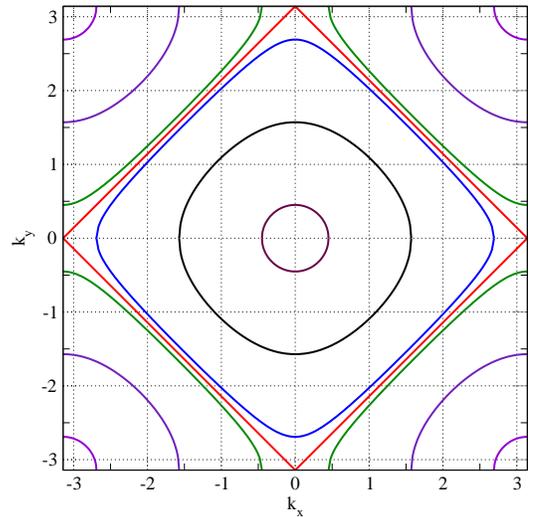}
    \caption{Constant energy contours for two-dimensional tight
      binding model on the square lattice in Eq. (\ref{energytb}).
The red square corresponds to the particle hole symmetric contour
where $\epsilon(\vec{k})=0$. The contours inside it are for negative $\epsilon(\vec{k})$
and those outside are for positive $\epsilon(\vec{k})$.
}
    \label{nrgtb}
  \end{center}
\end{figure}

Next, we will discuss scaling of the modulation length in with the chemical
potential, $\mu$ in the familiar tight binding models on the square and
triangular lattices at zero temperature.
%%%%%%%%%%%%%%%%%%%%%%%%%%%%%%%%%%%%%%%%%%%%%%%%%%
\subsubsecn{Tight binding model on the square lattice}
%%%%%%%%%%%%%%%%%%%%%%%%%%%%%%%%%%%%%%%%%%%%%%%%%%
We consider a two-dimensional tight binding model of the square lattice.
The dispersion in this model is given by
\begin{eqnarray}
\epsilon(\vec{k})=-2t\left( \cos k_x+\cos k_y  \right).
\label{energytb}
\end{eqnarray}
The constant energy contours corresponding to Eq. (\ref{energytb}) are drawn in Fig. \ref{nrgtb}.
As is clear from Fig. \ref{nrgtb}, there are certain directions (e.g., along the $X$-axis) along which there is no 
$\vec{k}$ for $\epsilon(\vec{k})>0$.
If we consider the same system at zero temperature, the following three crossovers are observed.\\
{\bf (i)} {\em Half filling:}
The chemical potential $\mu$ is zero at the half filling state.
The Fermi surface is given by $\pm k_x\pm k_y=\pi$. For small $\mu$, we have,
\begin{eqnarray}
\pm k_x\pm k_y=\pi+\frac{\mu}{2t\sin k_x},
\end{eqnarray}
thus giving us an uninteresting modulation exponent, $\upsilon_L=1$.\\
{\bf (ii)}  {\em Empty band:}
When $\mu=-4t$, none of the states are occupied. As we increase $\mu$ by a tiny amount $\delta\mu$ above this value, we observe 
a non-zero modulation wave-vector, $k=\sqrt{\delta\mu/t}$, thus showing a modulation exponent $\upsilon_L=1/2$.\\
{\bf (iii)} {\em Full inert bands:}
When $\mu=+4t$, all the states are occupied. As we lower $\mu$ by a tiny amount $\delta\mu$ below this value, we observe 
a difference $\delta k$ of the modulation vector from $\pm\hat{e}_x\pi\pm\hat{e}_y\pi$.
We have, $\delta k=\sqrt{\delta\mu/t}$, thus showing a modulation exponent $\upsilon_L=1/2$ again.

\begin{figure*}
  \begin{center}
%    \vspace{.2in}
%    \hspace{-1in}
\subfigure[ $~\mu=1.8$]{\includegraphics[width=.6\columnwidth]{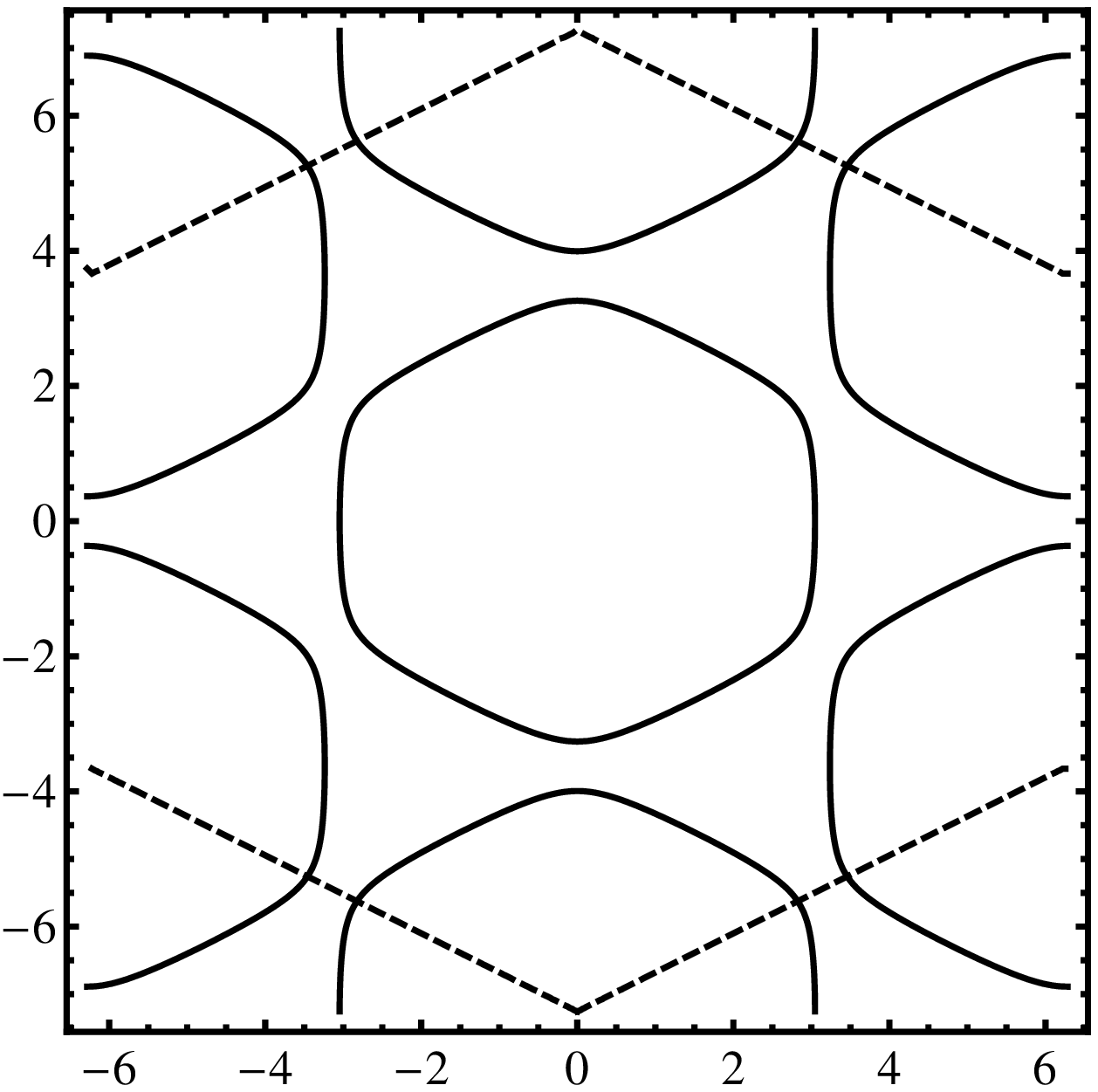}}\qquad
\subfigure[ $~\mu=2.2$]{\includegraphics[width=.6\columnwidth]{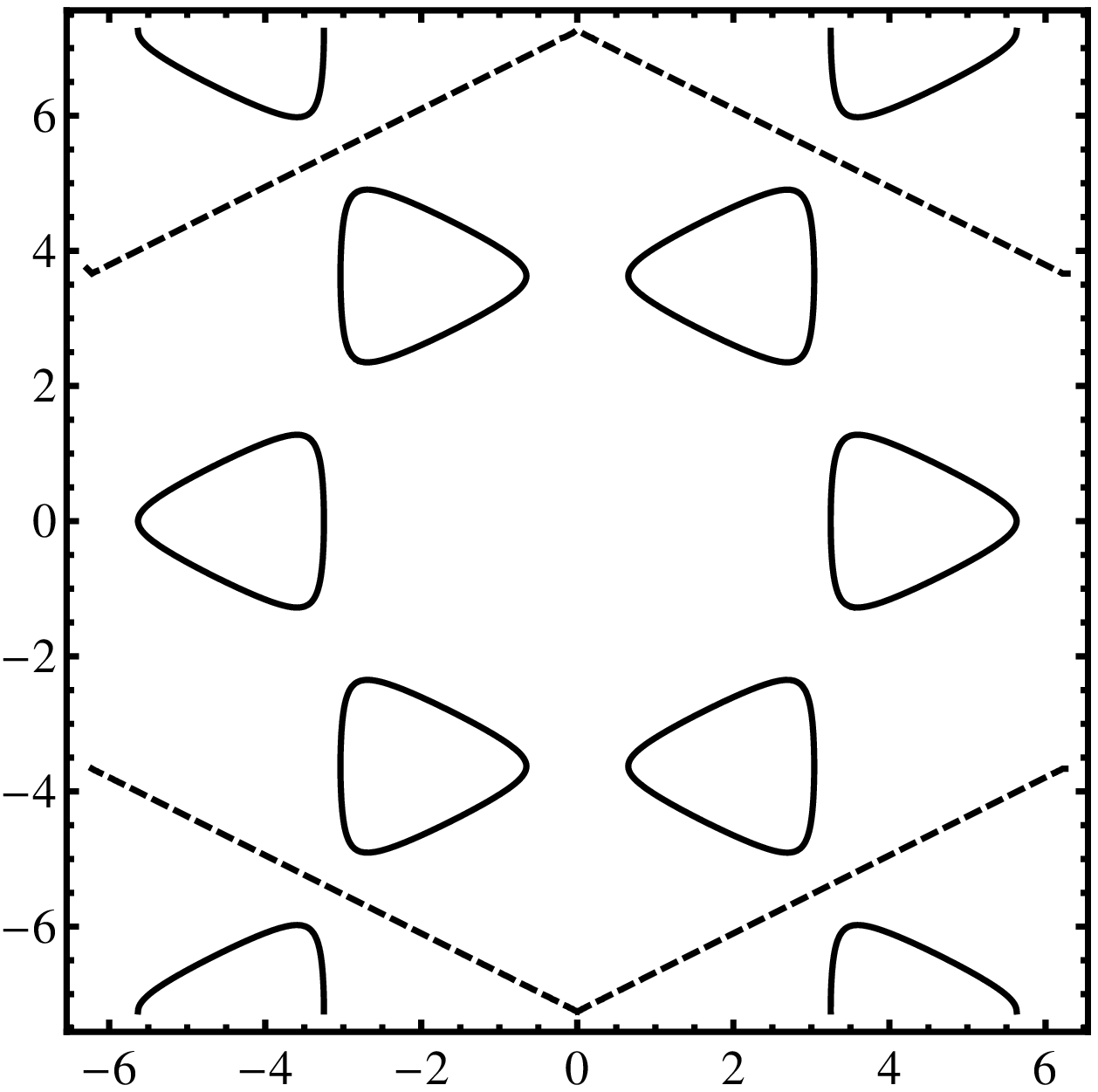}}\qquad
\subfigure[ $~\mu=2.2$(blue), $\mu=2$(green) and $\mu=1.8$(red)]{\includegraphics[width=.6\columnwidth]{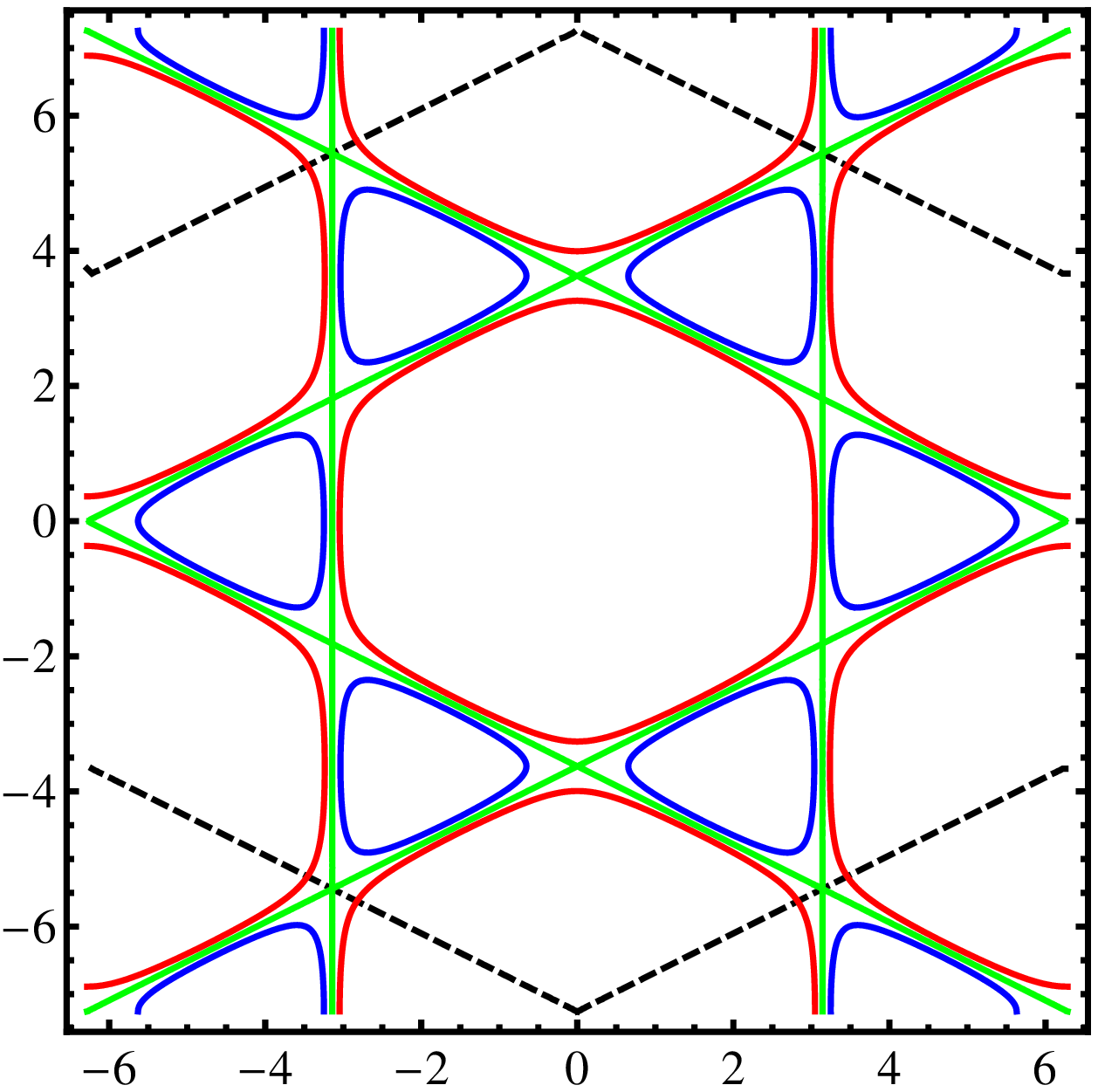}}
    \caption{Fermi surface for a triangular lattice with tight binding. The dashed lines are the Brillouin zone boundaries.
This demonstrates a smooth crossover from one set of Fermi surface branches to another as $\mu$ is changed across $\mu=2$.
The points where the crossovers take place are $(0,\pm 2\pi/\sqrt{3})$, $(\pm\pi,\pm\pi/\sqrt{3})$.
The modulation length exponent for this crossover is $\upsilon_L=1$.}
    \label{triangularfermi}
  \end{center}
\end{figure*}

%%%%%%%%%%%%%%%%%%%%%%%%%%%%%%%%%%%%%%%%%%%%%%%%%%
\subsubsecn{Tight binding model on the triangular lattice}
%%%%%%%%%%%%%%%%%%%%%%%%%%%%%%%%%%%%%%%%%%%%%%%%%%
The analysis of the triangular lattice within the tight binding approximation, is very
similar to the square lattice discussed above.
The dispersion $\epsilon(\vec{k})$ is given by
\begin{eqnarray}
\epsilon(k)=-2t\cos k_x-4t \cos \frac{k_x}{2}\cos \frac{k_y\sqrt{3}}{2}.
\end{eqnarray}
We have exponents similar to the square lattice.\\
{\bf (i)} {\em Three-quarters filling:}
The chemical potential $\mu=2t$ corresponds to the three-quarters filling state.
If we concentrate on the $\{k_x=\pi,~k_y:-\pi/\sqrt{3}\to\pi/\sqrt{3}\}$ segment (same phenomenon is present at
all the other segments of the quarter filling Fermi surface), we get,
\begin{eqnarray}
\delta k_x\sim\frac{\delta\mu}{2\cos\left(\frac{k_y\sqrt{3}}{2}\right)},
\end{eqnarray}
where $k_x=\pi+\delta k_x$ is obtained when $\mu=2t+\delta\mu$.
This leads to a modulation exponent of $\upsilon_L=1$.
The Fermi surfaces for chemical potentials $\mu$ close to three-quarters
filling are schematically shown in Fig. \ref{triangularfermi}. \\
{\bf (ii)}  {\em Empty band:}
When $\mu=-6t$, none of the states is occupied. As we increase $\mu$ by a tiny amount $\delta\mu$ above this value, we observe 
a non-zero modulation wave-vector, $k=\sqrt{2\delta\mu/3}$, thus showing a modulation exponent $\upsilon_L=1/2$.\\
{\bf (iii)} {\em Full inert bands:}
When $\mu=3t$, all of the states are occupied and close to this value
the Fermi surface is composed of six small circles around
$\vec{k}=\hat{x}\cos(n\pi/3)+\hat{y}\sin(n\pi/3)$,
$n=\{0,1,2,3,4,5\}$. If $\mu=3t-\delta\mu$, we get, 
$|\vec{\delta k}|=2\sqrt{\delta\mu/3}$, again giving us a modulation
length exponent, $\upsilon_L=1/2$.

%%%%%%%%%%%%%%%%%%%%%%%%%%%%%%%%%%%%%%%%%%%%%%%%%%
\subsubsecn{Metal-Insulator transition}
%%%%%%%%%%%%%%%%%%%%%%%%%%%%%%%%%%%%%%%%%%%%%%%%%%
We discuss here the metal to band insulator transition at zero temperature.
In a non-interacting system, this occurs
when the Fermi energy is changed such that all occupied bands become completely full,
as shown in Fig. \ref{metins}.
In the insulator, the 
% correlation length $\xi$ is finite, thus forbidding
% any long range effective hopping of fermions. 
% The 
Fermi energy lies in between two bands and thus the
filled states are separated from the empty states by a finite energy gap.
As the Fermi energy is tuned, the Fermi energy might touch one of the
bands thereby rendering the system metallic.
Close to this transition, the energy is quadratic in the momentum $k$, i.e.,
$|k|\propto|\delta\mu|^{1/2}$. 
This implies that,
\begin{eqnarray}
% |\mbox{ Re }\delta k|&\propto&|\delta\mu|^{1/2}\mbox{, and}\nonumber\\
% |\mbox{ Im }\delta k|&\propto&|\delta\mu|^{1/2}
|\delta k|&\propto&|\delta\mu|^{1/2}.
\end{eqnarray}
Following the scaling convention in Eq. (\ref{dle}), we adduce a similar exponent 
\begin{eqnarray}
\upsilon_L=1/2
\label{ups}
\end{eqnarray} that
governs the scaling of the modulation lengths with the shift $\delta \mu$ of the chemical potential
(instead of temperature variations).
% As usual, the exponent $\upsilon_L$
% characterizes changes in the chemical potential $\mu$ which lead to
% deviations of the modulation length from the value which appears at
% the point where the system turns metallic from the insulating regime.
% constant modulation length as $\mu$ crosses a fixed value $\mu_*$.

% As usual, the exponent $\upsilon_L$
% characterizes changes in the chemical potential $\mu$ which lead to a
% constant modulation length as $\mu$ crosses a fixed value $\mu_*$.
% The exponent $\upsilon_c$ here, represents divergences in the fermion hopping
% distance/the penetration depth as the system changes from an
% insulator to a metal.

%%%%%%%%%%%%%%%%%%%%%%%%%%%%%%%%%%%%%%%%%%%%%%%%%%%%%%%%%%%%
\subsubsecn{Dirac systems}% -- Graphene}
%%%%%%%%%%%%%%%%%%%%%%%%%%%%%%%%%%%%%%%%%%%%%%%%%%%%%%%%%%%%
The low energy physics of graphene and Dirac systems 
is characterized by the existence of Dirac points in momentum space 
where the density of states vanishes and the energy, $\epsilon(k)$
is proportional to the momentum $k$ for small $k$. When we invoke and repeat our
earlier analysis to these systems, we discern a trivial exponent
\begin{eqnarray}
|\delta k|&\propto&|\delta\mu|\nonumber\\
&\implies&\upsilon_{Dirac}=1.
\end{eqnarray}
This exponent may be contrasted with that derived from Eq. (\ref{ups}).

%%%%%%%%%%%%%%%%%%%%%%%%%%%%%%%%%%%%%%%%%%%%%%%%%%%%%
\subsubsecn{Topological Insulators -- Multiple length scale exponents as
  a function of the chemical potential $\mu$}
%%%%%%%%%%%%%%%%%%%%%%%%%%%%%%%%%%%%%%%%%%%%%%%%%%%%%                                     
\begin{figure*}
  \begin{center}
    \vspace{.2in}
\subfigure[]{\includegraphics[width=.6\columnwidth]{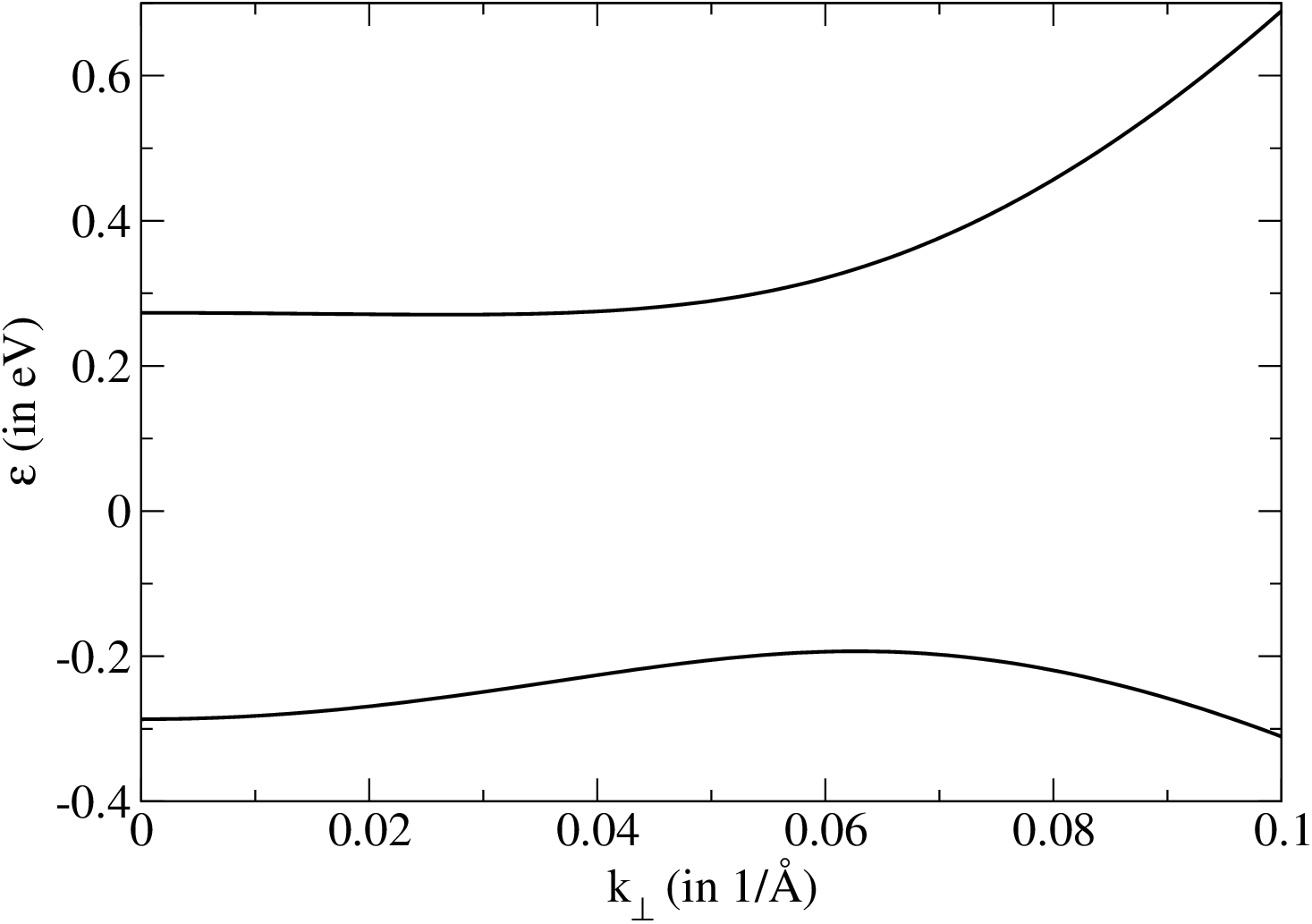}\label{etopinsperp}}\qquad
\hspace{.1in}
\subfigure[]{\includegraphics[width=.6\columnwidth]{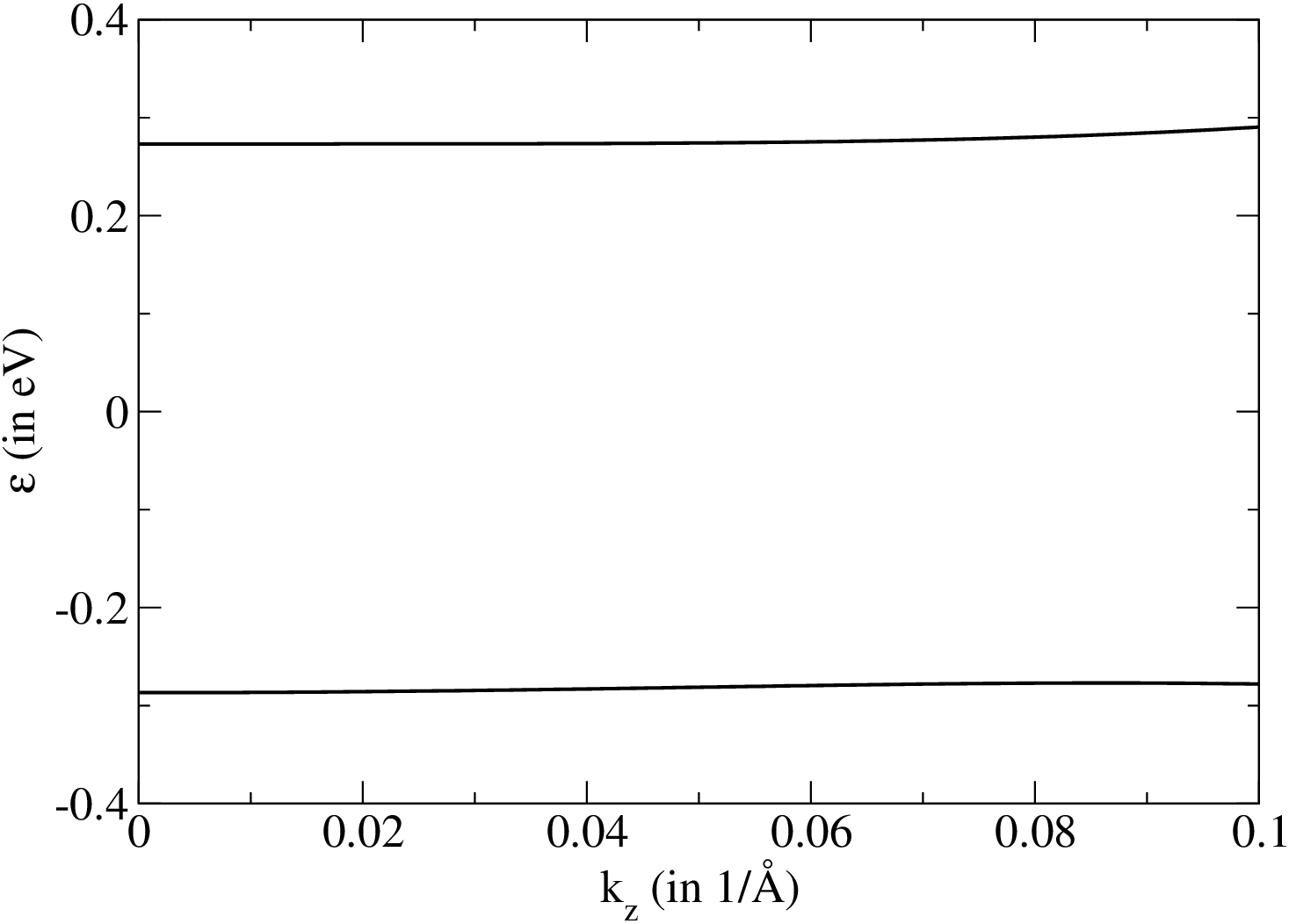}\label{etopinsz}}\qquad
\hspace{.1in}
\subfigure[]{\includegraphics[width=.6\columnwidth]{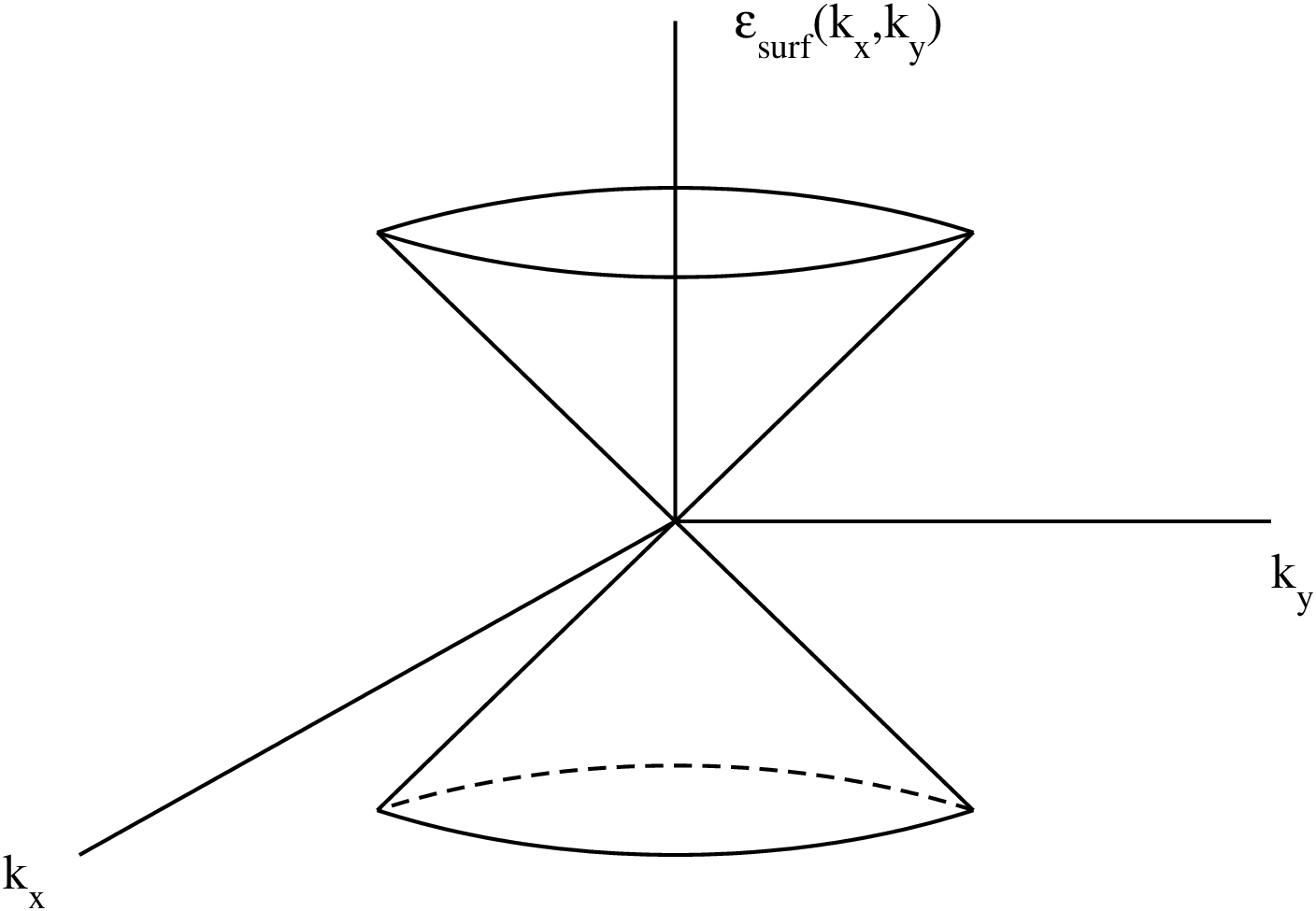}\label{esurftopins}}
    \caption{Energy levels of $Bi_2Se_3$ topological insulator.\\
\ref{etopinsperp}: $\epsilon(\vec{k})$ versus $k_{\perp}$ at $k_z=0$;
\ref{etopinsz}: $\epsilon(\vec{k})$ versus $k_z$ at $k_{\perp}=0$;
\ref{esurftopins}: $\epsilon_{surf}(k_x,k_y)$ versus $\vec{k_{\perp}}\equiv(k_x,k_y)$.
}
    \label{etopins}
  \end{center}
\end{figure*}

% \begin{figure}
%   \begin{center}
%     \vspace{.2in}
% \subfigure[]{\includegraphics[width=.9\columnwidth]{etopinsperp.eps}\label{etopinsperp}}\\
% \vspace{.2in}
% \subfigure[]{\includegraphics[width=.9\columnwidth]{etopinsz.eps}\label{etopinsz}}\\
% \vspace{.2in}
% \subfigure[]{\includegraphics[width=.9\columnwidth]{esurftopins.eps}\label{esurftopins}}
%     \caption{  
% \ref{etopinsperp}: $\epsilon(\vec{k})$ versus $k_{\perp}$ at $k_z=0$;
% \ref{etopinsz}: $\epsilon(\vec{k})$ versus $k_z$ at $k_{\perp}=0$;
% \ref{esurftopins}: $\epsilon_{surf}(k_x,k_y)$ versus $k_{\perp}$.}
%     \label{etopins}
%   \end{center}
% \end{figure}

The quintessential low energy physics of three-dimensional topological insulators can be gleaned from
the following effective Hamiltonian\cite{topinsham} in momentum space,
\begin{eqnarray}
H(\vec{k})&=&\epsilon_0(\vec{k})I_{4\times4}+\nonumber\\
&&\left(
\begin{array}{cccc}
{\cal M}(\vec{k}) & A_1k_z & 0 & A_2k_-\\
A_1k_z & -{\cal M}(\vec{k}) & A_2k_- & 0\\
0 & A_2k_+ & {\cal M}(\vec{k}) & -A_1k_z\\
A_2k_+ & 0 & -A_1k_z & -{\cal M}(\vec{k})
\end{array}
\right)
\end{eqnarray}
where 
$\epsilon_0(\vec{k})=C+D_1k_z^2+D_2k_{\perp}^2$,
${\cal M}(\vec{k})=M-B_1k_z^2-B_2k_{\perp}^2$ ,
with
$k_{\pm}=k_x+ik_y$,
$k_{\perp}=\sqrt{k_x^2+k_y^2}$ and
$A_1$, $A_2$, $B_1$, $B_2$, $C$, $D_1$ and $D_2$ constants for a given
system.
The energy bands are given by
\begin{eqnarray}
\epsilon(\vec{k})=\epsilon_0(\vec{k}) \pm \sqrt{{\cal M}(\vec{k})^2+A_1k_z^2+A_2k_{\perp}^2}.
\end{eqnarray}
These bands are plotted in Figs. \ref{etopinsperp} and \ref{etopinsz}.
The finite gap between the two bands leads to an exponentially damped
hopping amplitude, characterized by a finite correlation length when
the Fermi energy lies within this gap.
These energy bands disperse quadratically for small $k$ thus 
yielding 
\begin{eqnarray}
|\delta k|&\propto&\sqrt{|\delta\mu|}\nonumber\\
&\implies&\upsilon_{bulk}=1/2
\end{eqnarray}
whenever the correlation length diverges and a insulator to metal transition
takes place in the bulk, thus allowing long range hopping. The same exponent is also expected whenever
the modulation length becomes constant as $\mu$ crosses some threshold
value. 
% The surface states on the other hand
% are not always so smooth. For the same systems as above 

The effective
Hamiltonian for the surface states is given by
\begin{eqnarray}
H_{surf}=\left(
\begin{array}{cc}
0 & A_2k_-\\
A_2k_+ & 0
\end{array}
\right),
\end{eqnarray}
leading trivially to surface energies
\begin{eqnarray}
\epsilon_{surf}(k_x,k_y)=\pm A_2k_{\perp}.
\end{eqnarray}
% Thus the energy profile at the surface is like Fig. \ref{esurftopins}.
% Just like 
Similar to the Dirac points in graphene (see Fig. \ref{esurftopins}),
we trivially find an exponent of
\begin{eqnarray}
\upsilon_{surf}=1.
\end{eqnarray}

\subsubsecn{An example of a zero temperature Fermi system in which
  $\upsilon_L$ is not half or one}
Very large (or divergent) effective electronic masses $m_{eff}$ can be found 
in heavy fermion systems (and at putative quantum critical points).\cite{my_review,
  colemandiveffmassqcp}
If the electronic dispersion $\epsilon(\vec{k})$ has a minimum at $\vec{k}_{0}$ then a Taylor expansion about that minimum trivially reads
\begin{eqnarray}
\epsilon(\vec{k})=\epsilon(\vec{k}_0)+\frac{\hbar^2}{2}\sum_{ij}\left(m^{-1}_{eff}\right)_{ij}(k_i-k_{0i})(k_j-k_{0j})+\nonumber\\
\sum_{ijl}A_{ijl}(k_i-k_{0i})(k_j-k_{0j})(k_l-k_{0l})+ \ldots .
\end{eqnarray}
When present, parity relative to $\vec{k}_{0}$ or other considerations
may limit this expansion to contain only even terms. As an example, we
consider the dispersion
\begin{eqnarray}
\epsilon(k)=c_1-c_2(k^2-k_0^2)^4,
\end{eqnarray}
where $c_2>0$.
The hopping correlation function of such a system has a term which exhibits modulations
at wave-vector $k=k_0$ at $\mu=\mu_*=c_1$. At higher values of the
chemical potential, such a term ceases to exist.
At lower values ($\mu=\mu_*-\delta\mu$), this term breaks up into two terms whose modulation wave-vectors
are different from $k_0$ by,
\begin{eqnarray}
k-k_0&\sim&\pm\frac{\delta\mu^{1/4}}{2k_0c_2^{1/4}},\nonumber\\
&\implies&\upsilon_L=1/4.
\end{eqnarray}

%%%%%%%%%%%%%%%%%%%%%%%%%%%%%%%%%%%%%%%%%%%%%%%%%%
\subsecn{Finite temperature length scales -- Scaling as a
  function of temperature}
%%%%%%%%%%%%%%%%%%%%%%%%%%%%%%%%%%%%%%%%%%%%%%%%%%
At finite temperatures, apart from the modulation lengths,
%(as for the zero temperature situation), 
there generally is a set of characteristic correlation lengths.
From Eq. (\ref{gxfermi}), these are obtained by finding the poles (or
other singularities) of the
Fermi function. Along some direction $\hat{e}_0$, the wave-vector
$\vec{k}_0=\hat{e}_0k_0$ is associated with a pole
$k_0=\pm2\pi/L_{0}\pm i/\xi_0$. At this wave-vector,
\begin{eqnarray}
\epsilon(\vec{k_0})=\mu+\frac{2n+1}{\beta}i,
\label{poles_finite}
\end{eqnarray}
where $n$ is an integer.
For a given $\mu$, let us suppose that as we change the temperature, at $T=T_0$,
we reach a saddle point of $\epsilon(\vec{k})$ in the complex plane of
one of the Cartesian components of $\vec{k}$.
Then, near this saddle point, the corresponding correlation and modulation lengths scale as,
\begin{eqnarray}
|L_D-L_{D0}|&\propto&|T-T_0|^{\nu_L},\nonumber\\
|\xi-\xi_0|&\propto&|T-T_0|^{\nu_c},
\end{eqnarray}
where $\nu_L=\nu_c=1/2$ in most cases (when the second derivative is not zero).

%%%%%%%%%%%%%%%%%%%%%%%%%%%%%%%%%%%%%%%%%%%%%%%%%%%%%%%%%%%%
\secn{Euler-Lagrange equations for scalar spin systems}\label{appeullag}
%%%%%%%%%%%%%%%%%%%%%%%%%%%%%%%%%%%%%%%%%%%%%%%%%%%%%%%%%%%%
% In this appendix, we discuss an alternative approach for analyzing our systems.
%In this approach, we study 
We elaborate on the Euler-Lagrange equations associated with the free
energy of Eq. (\ref{freenrg}) in Sec. \ref{secchaos}.
% The free energy of our system in the continuum is given by
% \begin{eqnarray}
% {\cal
%   F}[s]=\frac{1}{2}\int\frac{d^dk}{(2\pi)^d}(v(\vec{k})+\mu)|s(\vec{k})|^2
% \end{eqnarray}
% Adding a quartic interaction, like in Eq. (\ref{Ham_soft}), we have,
% \begin{eqnarray}
% {\cal
%   F}[s]&=&\frac{1}{2}\int\frac{d^dk}{(2\pi)^d}(v(\vec{k})+\mu)|s(\vec{k})|^2+\nonumber\\
% &&~~~~~~~~~\frac{u}{4}\int
% d^dx (S^2(\vec{x})-1)^2.\label{freenrg}
% \end{eqnarray}
% This corresponds to an Ising system in the limit of large $u$.
% For finite or small $u$, this free energy corresponds to a
% ``soft-spin'' model in which $S(\vec{x})$ is not exactly normalized.
% We treat the free energy in Eq. (\ref{freenrg}) as an effective
% action.
% Variation of the field $S$ about the extremum of ${\cal F}$ gives us
% the following Euler-Lagrange equations,
These assume the form,
\begin{eqnarray}
\int d^dy \tilde{V}(\vec{x}-\vec{y}) S(\vec{y})+\mu S(\vec{x})\nonumber\\
+u(S^2(\vec{x})-1)S(\vec{x})=0,\label{eullag}
\end{eqnarray}
where $\tilde{V}(\vec{x})=[V(\vec{x})+V(-\vec{x})]/2$.
For example, if we consider the finite ranged system for which,
\begin{eqnarray} 
\int d^dy \tilde{V}(\vec{x}-\vec{y}) S(\vec{y})&=&a\nabla^2
S(\vec{x})\nonumber\\
&&~+b\nabla^4 S(\vec{x})+ \ldots,
\label{abdelfinite}
\end{eqnarray}
then, we have,
\begin{eqnarray}
a\nabla^2 S(\vec{x})&+&b\nabla^4 S(\vec{x})+ \ldots
+\mu S(\vec{x})\nonumber\\
&+&u(S^2(\vec{x})-1)S(\vec{x})=0.
\end{eqnarray}

For lattice systems, the Euler Lagrange equation (\ref{eullag}) 
reads
\begin{eqnarray}
\sum_{\vec{y}} \tilde{V}(\vec{x}-\vec{y}) S(\vec{y})&+&\mu
S(\vec{x})\nonumber\\
&+&u(S^2(\vec{x})-1)S(\vec{x})=0.
\end{eqnarray}
In general, it may be convenient to express the linear terms in the
above equation in terms of the
lattice Laplacian $\Delta$.
We write
\begin{eqnarray}
D(\Delta) S(\vec{x})\equiv\sum_{\vec{y}} \tilde{V}(\vec{x}-\vec{y})
S(\vec{y})+\mu S(\vec{x}),
\end{eqnarray}
$D$ being some operator which is a function of the lattice Laplacian
$\Delta$.
The real-space lattice Laplacian $\Delta$, given by the Fourier
transform of Eq. (\ref{LLaplace}), acts on a general field $f$ as
\begin{equation}
\Delta f(\vec{x})\equiv-\sum_{i=1}^d [f(\vec{x}+\hat{e}_i)+f(\vec{x}-\hat{e}_i)-2f(\vec{x})].
\end{equation}
Here, $\{\hat{e}_i\}$ denote unit vectors along the Cartesian directions.
(In the continuum limit, $\Delta$ can be replaced by $-\nabla^2$.)
The Euler-Lagrange equation then, takes the form,
\begin{eqnarray}
D(\Delta) S(\vec{x})+u(S^2(\vec{x})-1)S(\vec{x})=0.\label{eullaglat}
\end{eqnarray}

Equation \ref{abdelfinite} corresponds, on the lattice, to
% \begin{eqnarray} 
% \sum_{\vec{y}} \tilde{V}(\vec{x}-\vec{y}) S(\vec{y})=-a\Delta S(\vec{x})+b\Delta^2 S(\vec{x})+~.~.~.~~.
% \end{eqnarray}
\begin{eqnarray} 
\sum_{\vec{y}} \tilde{V}(\vec{x}-\vec{y}) S(\vec{y})&&=\nonumber\\
&&-a\Delta S(\vec{x})+b\Delta^2 S(\vec{x})+ \ldots .
\end{eqnarray}
The Euler Lagrange equation for this finite ranged system reads
\begin{eqnarray}
-a\Delta S(\vec{x})&+&b\Delta^2 S(\vec{x})+ \ldots
+\mu S(\vec{x})\nonumber\\
&+&u(S^2(\vec{x})-1)S(\vec{x})=0.
\end{eqnarray}

\end{appendix}

\bibliography{$HOME/Dropbox/docs/mybiblio}

\begin{thebibliography}{85}
\bibitem[{mze()}]{mzerononzero}
\bibinfo{note}{The order of the derivative $m$ in Eq. (\ref{lengthsK}) is zero
  if $G(k)$ has a pole of finite order at $k=K$, and $m\ge0$ for the branch
  points.}
\bibitem[{not()}]{noteentire}
\bibinfo{note}{If we do not have any pole or branch point of the Fourier space
  correlation function, the form of the real space correlation function is
  governed by the endpoints of the $k$-space integration. Therefore, in the
  continuum, no finite modulation length is allowed. The same is true for the
  correlation length.}
\bibitem[{ons()}]{onsystem}
\bibinfo{note}{An $O(n)$ system is one for which the order parameter
  $\vec{S}(\vec{x})$ has $n$ components normalized as
  $\vec{S}(\vec{x})\cdot\vec{S}(\vec{x})=n$. The large $n$ limit of this system
  is equivalent to the spherical model where the spins are constrained only by
  the relation $\sum_{\vec{x}}(S(\vec{x}))^2=N$, where $N$ is the number of
  lattice sites. See H. E. Stanley, Phys. Rev. {\bf 176}, 2, 718 (1968).}
\end{thebibliography}

\end{document}